\documentclass{article}

\usepackage{arxiv}

\usepackage[utf8]{inputenc} % allow utf-8 input
\usepackage[T1]{fontenc}    % use 8-bit T1 fonts
\usepackage{hyperref}       % hyperlinks
\usepackage{url}            % simple URL typesetting
\usepackage{booktabs}       % professional-quality tables
\usepackage{amsfonts}       % blackboard math symbols
\usepackage{nicefrac}       % compact symbols for 1/2, etc.
\usepackage{microtype}      % microtypography
\usepackage{makecell}
\usepackage{siunitx}
\usepackage{caption}
\usepackage{subcaption}
\usepackage{multirow}
\usepackage{xr}
\usepackage{cleveref}
\usepackage{graphicx}
\usepackage{enumitem}

\pdfoutput=1

\title{Beyond Words: An Experimental Study of Signaling in Crowdfunding}

\author{
 Henry K. Dambanemuya \\
  Northwestern University\\
  Evanston, IL 60208 \\
  \texttt{hdambane@u.northwestern.edu} \\
   \And
  Eunseo Choi \\
  MIT\\
  Cambridge, MA 02139 \\
  \texttt{choie@mit.edu} \\
   \And
  Darren Gergle \\
  Northwestern University\\
  Evanston, IL 60208 \\
  \texttt{dgergle@northwestern.edu} \\
   \And
  Em\H oke-\'Agnes Horv\'at \\
  Northwestern University\\
  Evanston, IL 60208 \\
  \texttt{a-horvat@northwestern.edu} \\
}

\begin{document}
\maketitle

\begin{abstract}
Increasingly, crowdfunding is transforming financing for many people worldwide. Yet we know relatively little about how, why, and when funding outcomes are impacted by signaling between funders. We conduct two studies of $N=500$ and $N=750$ participants involved in crowdfunding to investigate the effect of certain characteristics of ``crowd signals'' on the decision to fund. We find that, under a variety of conditions, contributions of heterogeneous amounts arriving at varying time intervals are significantly more likely to be selected than homogeneous contribution amounts and times. The impact of signaling is strongest among participants who are susceptible to social influence. The effect is remarkably general across different project types, fundraising goals, participant interest in the projects, and participants' altruistic attitudes. Critically, the role of crowd signals in decision-making is typically unrecognized by participants. Our results underscore the fundamental nature of social signaling in crowdfunding, informing strategies for platforms, funders, and project creators.
\end{abstract}

\keywords{crowdfunding, crowdsourcing, crowd signals, social influence, altruistic attitudes}

\section{Introduction}

Online fundraising (aka ``crowdfunding'') is the process of funding a project or venture by raising small amounts of money from many people outside of traditional financial institutions, typically via Web-based platforms~\cite{mollick2014dynamics}. It is a rapidly-growing industry with several applications, including disaster relief, political campaigns, and the support of personal, artistic, entrepreneurial, and scientific endeavors~\cite{vachelard2016guide,kuo2012design,vulkan2016equity,lee2022new}. Due to its broad societal relevance, crowdfunding has attracted significant interest in industry and policy-making~\cite{whitehouse2016} and extensive research in various fields including social computing, entrepreneurship, law, public management, and the social sciences~\cite{zhang2018bibliometric}. Most of this work has focused on discovering factors that differentiate highly visible success stories (e.g., Khushi Baby\footnote{\url{www.khushibaby.org}}, Oculus Rift\footnote{\url{www.oculus.com/rift}}, or the Ocean Cleanup\footnote{\url{www.theoceancleanup.com}}) from the many crowdfunding projects that fail to reach their target amount~\cite{mollick2014dynamics,greenberg_learning_2014}\footnote{For example, between 2014 and 2022, only 38\% of projects posted on Kickstarter and 13.3\% of projects on Indiegogo were successfully funded~\cite{crowddata2022}.}.

Recent work in human-computer interaction (HCI) and design investigates social translucence in crowdfunding communities, establishing the importance of making information visible to potential contributors\footnote{Crowdfunding literature has used various terms to denote people who pledge funds on various types of platforms, calling them ``investors,'' ``donors,'' ``backers,'' or ``supporters.'' Here, we refer to them as ``contributors'' or ``funders.''}~\cite{kim2020enriched}. Examining crowdfunding in a social translucence framework adds to an extensive HCI record that aims to improve visibility, awareness, and accountability on digital platforms ranging from Wikipedia~\cite{suh2008lifting} to social networks~\cite{gilbert2012designing} and explainable AI~\cite{ehsan2021expanding}. 
Social translucence theory argues that online collaboration systems facilitate the attainment of common goals by making users' activities visible to one another~\cite{erickson2000social,erickson2002social}. 

Indeed, when deciding which projects to fund, potential contributors on crowdfunding platforms pay attention to the characteristics of projects and their creators, such as linguistic features~\cite{larrimore2011peer,althoff2014ask,mitra2014language,rhue_emotional_2018,gafni2019life,anglin2018power}, human capital~\cite{colombo2015internal,greiner2009role,zheng2014role}, project updates~\cite{xu2014show} and engaging media content~\cite{dey2017art}. Thus, awareness of information about projects on crowdfunding platforms is a key first step in helping align users' decisions with their priorities. 
However, despite the visibility of various project details, there is an information asymmetry between creators and their funders, who are risking their own funds and time looking for meritorious projects. Prospective funders observe other funders' contributions on the platform to counter this asymmetry. For instance, they inspect how much money a project received and when, weighing clues about the right amount and time to contribute~\cite{shang2009field,zhang2012rational, ceyhan2011dynamics,vulkan2016equity,burtch2013empirical,dambanemuya2019harnessing,van2020follow,horvat2023hidden}.

Despite this common practice in crowdfunding, we know relatively little about the signaling encoded in prior contributions that might sway potential funders who are on the verge of funding a project. Arguably, this represents one of the most puzzling open problems related to crowdfunding. Prior work indicates that signaling is more likely to occur when it is difficult to establish the merit of a project based on its description~\cite{spence2002signaling,salganik2006experimental,zhang2012rational,connelly2011signaling}. Signaling, however, leads to substantial herding~\cite{zhang2012rational,vismara2018information,astebro2019herding,zaggl2019small,chan2020bellwether}. Herding can have catastrophic or optimal outcomes, depending on whether contributors imitate random novices or serial funders with a successful track record~\cite{zakhlebin2019investor,dambanemuya2023herding}. The importance of signaling via prior contributions and signaling's downstream effects on funders' decision-making, project outcomes, and platform longevity prompt us to investigate \textit{how, when, and why funders act upon social signaling}.

Recent research based on large-scale observational data from real crowdfunding sites identifies signals associated with successful funding across different types of crowdfunding markets~\cite{dambanemuya2021multi}. These so-called \emph{crowd signals} are deduced from funding dynamics based on variations in the amounts and timing of prior contributions. Critically, \emph{crowd signals are more predictive than project-related factors of which projects will raise their target}~\cite{dambanemuya2021multi}. While evidence about the effectiveness of crowd signals in observational data obtained from complex, evolving online platforms is compelling, these platforms may contain confounding factors that interfere with the effect of the signals~\cite{chakraborty2019impact,thalerchoice,aragon2017detecting}. For example, quasi-causal techniques cannot handle differences in the available projects, elements of platform design, and funder resources. This shortcoming and the opportunity to learn more about funder's motivations, predispositions, and reasoning call for \emph{experiments that unpack the causal link between social signaling and individual decision-making}.

Here, we conduct randomized experiments that allow us to establish a causal link between prior contribution scenarios and individuals' decisions to fund a project under various conditions. We translate empirically derived crowd signals related to the amounts and timing of contributions into mock funding scenarios and ask study participants to choose which scenarios would make them more likely to contribute funds. We also collect and analyze participants' qualitative rationales behind their choices to uncover the drivers of their decision-making and to understand what signals they perceive. In a first study of $N=500$ subjects familiar with crowdfunding, we test the fundamental influence of crowd signals on potential contributors given various project-related factors (e.g., topic, requested amount, length and quality of the request) and different personality types and predispositions (e.g., participants' interest in the topic, their altruism, and their susceptibility to social influence). 
In a more realistic setting, our second study of $N=750$ participants scrutinizes if the crowd signals' influence holds in the context of competing project descriptions. In this study, we show participants two similar projects with different crowd signals, establishing whether less salient social signaling (obscured by different project descriptions) affects the impact of crowd signals. Not only does this study verify our results in the presence of multiple stimuli and conditions, but to our surprise, participants' reflections indicate that most people think project descriptions guide their decisions. Given our controls for multiple stimuli and the performed counterbalancing in terms of the order of projects and different versions of crowd signals, this could not have been the case. 

Our work contributes experimental designs with high external validity and allows us to understand how, when, and why signaling via prior contributions is linked to individual decision-making. To our knowledge, this research is the first to produce causal evidence regarding the impact of manipulating specific social signals. It strongly supports the positive effect of heterogeneous contribution amounts and times on the decision to fund. Our studies probe this effect across various factors related to the request itself, users' personalities, and altruistic attitudes. %, and demographic characteristics, such as age, gender, and socio-economic status. 
Theoretically, our work extends research on the implications of social translucence, adding to recent discussions about this theory in the context of crowdfunding communities~\cite{kim2020enriched}. Practically, our findings are of potential interest to platform designers and users wishing to harness crowd signals to improve information acquisition, resource allocation, and the long-term success of crowdfunding platforms. Since design choices govern what social signals are represented on crowdfunding platforms, it is imperative that we better understand signaling mechanisms that are influencing our decision-making with or without our knowledge.

\section{Related Literature and Development of Research Questions}
\label{related-work}
This section introduces prior work that informed our research questions and study designs. First, we briefly review the extensive literature on project-related correlates of fundraising success. Importantly, we describe crowd signals that have been shown empirically to be more effective than project-based characteristics in predicting who gets funding. This strand of observational work prompted us to investigate experimentally \emph{how} signaling works on crowdfunding sites via specific crowd signals. Second, we discuss relevant research on factors related to participants' motivation and predisposition, which could influence who is more susceptible to crowd signals. This line of work forms the basis for our research question on \emph{when} social signaling is most likely to influence individual decision-making. Third, we review existing work on reasons to choose specific crowdfunding projects over others. Gaps in this body of literature triggered our interest in uncovering \emph{why} signaling works, with a focus on whether people recognized the signals or not. We conclude this section with our research questions and a summary of how we translated them into our study designs.

\subsection{Moving from Project-related Features to Crowd Signals as Indicators of Success: How Does Signaling Work?}
\label{sec:RW-indicators}

Extensive research uncovers the factors associated with higher probabilities of successful fundraising. The majority of this work focuses on features of the projects, such as the project description~\cite{gafni2019life,althoff2014ask,larrimore2011peer,mitra2014language,rhue_emotional_2018}, attention-capturing media content~\cite{mitra2014language,mollick2014dynamics,dey2017art}, project updates~\cite{xu2014show,lee_content-based_2018}, promotional activities on social media~\cite{etter2013launch,lu2014inferring,zhang_predicting_2017,lu2014inferring}, the requested amount~\cite{mollick2014dynamics,cumming2015crowdfunding,zheng2014role,cordova2015determinants}, and the duration of the fundraising effort~\cite{cumming2015crowdfunding,cordova2015determinants}. Other studies have also looked at the characteristics of the project creators, such as their reputation~\cite{collier2010sending} and social capital~\cite{ahlers2015signaling,horvat2015network,vismara2016equity,zheng2014role,greiner2009role,chung2015long}. 

A parallel thread of work connects crowdfunding outcomes to signals deduced from the behavior of the contributing crowd. Extensive research demonstrates that the amount of the first contribution~\cite{koning2013experimental}, the timing of contributions~\cite{solomon2015don,colombo2015internal,agrawal2015crowdfunding}, and other descriptors of crowd dynamics~\cite{ceyhan2011dynamics,burtch2013empirical,agrawal2015crowdfunding} are correlated with fundraising success. For instance, having sizable initial amounts and many early contributors to a fundraising campaign may signal project quality and funders' confidence in a project -- factors that can ultimately lead to a project's success~\cite{colombo2015internal,koning2013experimental}. Additionally, having many early contributors can result in more opportunities to obtain subsequent contributions through potential information cascades and social influence~\cite{vismara2016information,zhang2012rational}. 

\begin{figure}[!h]
    \centering
    \includegraphics[scale=.45]{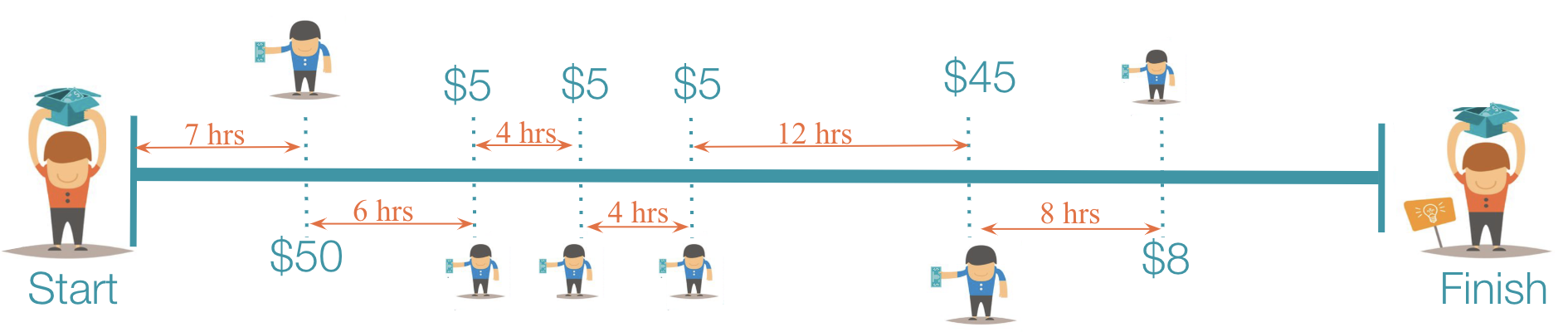}
    \caption{An illustration of crowd signals on an example project that receives six contributions. For example, the first contribution of \$50 arrives seven hours after the start of the campaign. In this example, the coefficient of variation in inter-contribution times is 0.44. The coefficient of variation in contribution amounts is 1.10. Our studies vary the contribution amounts and contribution timings, examining the effect of their variation on participants' decisions to contribute funds to mock crowdfunding projects.}
    \label{fig:illustration}
\end{figure}

Most importantly, recent work on multiple crowdfunding markets shows that signals from funding dynamics are more effective in determining the success of the fundraising effort than characteristics of the projects and their creators~\cite{dambanemuya2021multi}. These signals are the coefficient of variation in the inter-contribution times and the coefficient of variation in the contribution amounts. The coefficient of variation is computed as the standard deviation divided by the mean of the inter-contribution times and contribution amounts, respectively. Figure~\ref{fig:illustration} illustrates these two signals on an example project. 
Successfully funded projects are associated with greater variation in contribution amounts \textit{and} inter-contribution times compared to failed projects~\cite{dambanemuya2021multi}. Recent empirical work on large-scale peer-to-peer lending data shows that two similar signals, the Gini coefficient of contribution amounts and the average inter-contribution time, are \emph{jointly} predictive of loan repayment~\cite{horvat2023hidden}. Thus, a growing body of recent work suggests that aspects of the amounts and timing of contributions considered together are important predictors of fundraising success~\cite{dambanemuya2021multi,horvat2023hidden,solomon2015don}. Based on this prior evidence, we investigate the two signals on amounts and times together to validate their joint effect under a wide range of conditions.

The promise of these crowd signals is substantial. First, they can provide helpful heuristics for potential contributors when deciding whether to fund or not, even when data about projects and their creators is incomplete~\cite{ahlers2015signaling,vismara2018signaling}. Second, crowd signals capitalize on the promise of the wisdom of crowds~\cite{dambanemuya2019harnessing,dambanemuya2023herding}. Many active contributors on crowdfunding platforms have considerable experience and expertise as investors and can guide an attentive crowd to meritorious projects. Third, crowd behavior might provide early-warning signals through changes in contribution dynamics that reflect excitement and trust or conversely, uncertainty and disbelief~\cite{colombo2015internal,solomon2015don,horvat2023hidden}. Fourth, crowd signals can potentially provide an alternative to relying on traditional proxies of merit and creditworthiness that are prone to social biases~\cite{dobbie2021measuring}. Finally, differential crowd signals offer a potential explanation for how crowdfunding projects of similar quality can end up with different funding outcomes, depending on the collective approval conveyed via social signaling~\cite{stevenson2019out}. 

However, the role of the two crowd signals (the variation in contribution amounts and timing) has yet to be validated in a controlled setting. Such validation would enable us to deduce a direct link between crowd signals and individual decision-making, improving our understanding of \emph{how, when, and why} crowd signals work. This paper aims to fill this important gap in the literature.

\subsection{Motivations and Predispositions Impacting Behavior in Crowdfunding: When Does Signaling Work?} 
\label{sec:additionalfactors}
The literature on crowdfunding identifies several social factors that affect participation in and contributions on crowdfunding platforms. These factors include people's interest in the project topic (hereafter project category), as individuals prefer to support causes they are interested in personally~\cite{gerber_crowdfunding_2013,kraut2011encouraging}. For example, in online political crowdfunding, people's interest in politics affects participation~\cite{kusumarani2019people}. Additionally, the altruism of contributors increases their intention to participate, impacting the likelihood of a project's success~\cite{steigenberger2017supporters,giudici2018reward,rodriguez2019altruism}. Finally, contributors' susceptibility to social influence and compliance to social norms also influences their behavior in online fundraising~\cite{horvat2015network,hui2014understanding,liu2021social,kuppuswamy2018crowdfunding}.

While these studies provide ample evidence on how individuals' interest in the project topic, their altruism, and susceptibility to social influence affect their behavior in crowdfunding, less is known about how these factors may interact with crowd signals in determining \emph{when} people are more likely to contribute to a project. Therefore, our work also examines whether the effect of crowd signals depends on people's interest in a project, their altruism and susceptibility to social influence.

\subsection{The Effects of Social Signaling on Contributors' Behavior in Crowdfunding: Why Does Signaling Work?} 
\label{sec:RW-theoreticalPerspectives}
To uncover the motives behind people's responses to crowd signals, our study builds on literature identifying several reasons people get involved in crowdfunding campaigns~\cite{gerber_crowdfunding_2013,hui2014understanding}. These reasons include the need to help others, support a cause, be part of a community, and/or collect rewards. 

Prior studies also provide insights into how potential contributors decide on what projects to fund, persuaded by project characteristics or guided by crowd signals. Typically, potential funders observe the total amount already raised. Additionally, they pay attention to when and how much money other individuals have contributed.
A prominent human-computer interaction theory, \emph{social translucence theory}~\cite{erickson2000social}, suggests that the visibility of social information generates better coordination among funders to reach the fundraising goal. \emph{Signaling theory}~\cite{spence1973job,spence2002signaling} has also been widely used to explain why funders select certain projects over others. According to this theory, funders infer specific quality signals from project-creator and project-related indicators of success to inform their decision. While much is known about signaling through project-related information like amount requested, project description, or media content (e.g., ~\cite{reichenbach2021signals,vismara2018signaling,ahlers2015signaling,plummer2016better}), far less is understood about signaling through social information, i.e., how does observing others' prior contributions (the crowd signals) affect funders' behavior and why? 

On the one hand, \emph{conditional cooperation theory} suggests that individuals will contribute to projects that have received the most prior support~\cite{fischbacher2001people}. On the other hand, the \emph{underdog effect} suggests a higher tendency to contribute to projects that have received the least support~\cite{bradley2019relatively,zamudio2018e13}. Duncan's (2004) \emph{theory of impact philanthropy} further indicates that people prefer contributing to lesser funded projects because they derive satisfaction from knowing that their contribution will make a real difference to the fundraising effort~\cite{duncan2004theory}. Informed by these theoretical perspectives, we employ qualitative methods to investigate \emph{why} crowdfunding contributors choose projects with specific crowd signals.

\subsection{Research Questions}
Building on these threads of literature, we develop an experimental setup to systematically investigate how, when, and why social signaling works on crowdfunding platforms. Specifically, we design two studies that ask participants to select a project setting that they would contribute to. The project settings differ in the crowd signals and are developed such that we can control for various project-related characteristics.

\emph{Study I} examines the fundamental signaling mechanism emerging as potential funders inspect prior contribution to the \emph{same} project. This setting encourages people to focus on differences in two hypothetical contribution scenarios to the same project and asks the following research questions:

\begin{itemize}
    \setlength{\itemindent}{0.5em}
    \item[\textbf{RQ1}\emph{I}] Are crowd signals that encode high variation in the amounts and timing of contributions effective predictors of participants' selections? This question helps us understand \emph{how} crowd signaling works.
    \item[\textbf{RQ2}\emph{I}] Are crowd signals impacting decision-making under a suite of different stimuli? This question enables us to test the main effect in the presence of various project- and participant-related controls. Our controls include project category, fundraising goal, project description length and quality, people's interest level, altruistic attitudes, and susceptibility to social influence. This question inspects \emph{when} crowd signals work.
    \item[\textbf{RQ3}\emph{I}] Are participants aware of the effect of crowd signals on them? This question helps us unpack \emph{why} people make the selections they do by investigating their reasoning throughout the study.
\end{itemize}

\emph{Study II} expands Study I by investigating signaling in the context of two competing crowdfunding pitches. It asks the same questions centered on the how, when, and why of signaling with adjustments to the case of two crowdfunding pitches and their contributions to create a more realistic experimental setting:
\begin{itemize}
    \setlength{\itemindent}{0.5em}
    \item[\textbf{RQ1}\emph{II}] Are crowd signals that encode high variation in the amounts and timing of contributions effective predictors of participants' selections, even in the presence of two different project descriptions? This question furthers our knowledge of \emph{how} crowd signals work, elucidating the robustness of their effect in an inconspicuous setting.
    \item[\textbf{RQ2}\emph{II}] Are crowd signals impacting decision-making under a suite of different stimuli, including competing project pitches? This question further probes the universality of signaling, expanding our understanding of \emph{when} signals are effective.
    \item[\textbf{RQ3}\emph{II}] Are participants aware of the effect of crowd signals on them in the presence of different project pitches? Here, we address our \emph{why} question by identifying potential blind spots in people's awareness of social signaling.
\end{itemize}

\section{Study I: Fundamental Signaling Mechanism: The Single-Description Layout} 
In the first study, we investigate whether high variation in contribution amounts and timing makes people more likely to contribute to a crowdfunding project. Essentially, study participants see the description of a crowdfunding project and two hypothetical signaling scenarios that encode different crowd signals. Their task is to choose the signaling scenario that would make them more likely to contribute to the project. The single-description layout ensures that we test only for the effect of the manipulated crowd signals while minimizing the impact of any potential confounding variables. 

\subsection{Measures}
\label{sec: measures}
To establish a causal link between participants' choices (outcome) and crowd signals (predictors), we evaluate their association across a suite of conditions (controls) that enable us to move beyond a single-stimuli study. In what follows, we describe our outcome, predictor, and control variables.

\subsubsection{Outcome}
The outcome we seek to investigate is the number of times participants select a project under the two treatment conditions: high vs low variation conditions. To compute the outcome, we count participants' \emph{number of selections} in both treatment conditions.

\subsubsection{Predictors: The High vs Low Variation Conditions} 
Crowd signals are our predictor variables, and we map them to the two treatment conditions. The high variation condition shows contribution amounts and timing with high coefficients of variation. Prior work found that high variation in amounts and timing meant that projects were more likely to receive funding~\cite{dambanemuya2019harnessing,dambanemuya2021multi,horvat2023hidden}. Conversely, the low variation condition entails contribution amounts and timing with a low coefficient of variation. These have been associated with projects that are less likely to collect their target amount. To choose specific amounts and contribution times for high and low variation, we analyze data from prior observational studies that comprise nearly four million contributions from three different crowdfunding platforms~\cite{dambanemuya2021multi}. We first compute the average coefficient of variation in amounts and inter-contribution times on all platforms using the first four contributions for each project. We aim to create artificial crowd signals above the computed averages on all three real-world platforms for the high variation condition; crowd signals below average are used for the low variation condition. 

Specifically, to select the individual values for each contribution list, we consider realistic and easily understandable numbers for the contribution amounts (Table~\ref{tab:artficial-signals}). For example, a contribution list consisting of the amounts \$100, \$10, \$40, and \$ 250 has a variation in amounts of 1.068, which is higher than average based on observational data (average = $0.533$). It is also substantially larger than the variation of the list \$85, \$100, \$120, \$95 (variation in amounts = 0.147). Similarly, four contributions arriving two days, three hours, two hours, and one hour before the participant's selection indicate high variation in contribution times (1.833) compared to the average from observational data (average = $0.855$). Conversely, contributions arriving two days, 1.5 days, one day, and six hours before selection indicate low variation in times (0.659). Table~\ref{tab:artficial-signals} contains these values and another set of amounts and times that are computed based on the same logic.

We combine information about amounts and times for each treatment condition, ensuring that both variations are either high or low. This decision was motivated by prior work that points to the joint value of both variations in time and amount~\cite{dambanemuya2019harnessing,dambanemuya2021multi,horvat2023hidden}. To ensure that the results are robust to the exact choices of amounts and times, we create two sets of low and high values and combine them into four high and four low variation conditions. 

\begin{table}[!h]
\caption{Contribution amounts and times for high and low artificial crowd signals used in the treatment conditions. The contribution times shown to participants are obtained by subtracting each value from 48 hours as if the project would have started two days before the participant decides. To ensure that the contribution lists truly reflect high vs low variation in contribution amounts and timing, we use the same number of recent contributions (four), total amount raised (\$400), and time frame (48 hours).}
\label{tab:artficial-signals}
\centering
\begin{tabular}{|l|l|l|}
\cline{2-3}
 \multicolumn{1}{l|}{} & Condition (Value) & Artificial Values\\ \hline
Variation in  amounts & High A (1.068) & \$100, \$10, \$40, \$250 \\
 & High B (1.173) & \$85, \$15, \$30, \$270 \\ \cline{2-3} 
 & Low A (0.147) & \$85, \$100, \$120, \$95 \\
 & Low B (0.183) & \$80, \$110, \$120, \$90 \\ \hline
Variation in timing & High A (1.833) & 0hrs, 45hrs, 46hrs, 47hrs \\
 & High B (1.568) & 0hrs, 40hrs, 46hrs, 47hrs \\ \cline{2-3} 
 & Low A (0.659) & 0hrs, 12hrs, 24hrs, 42hrs \\
 & Low B (0.673) & 12hrs, 24hrs, 32hrs, 47hrs \\ \hline
\end{tabular}
\end{table}

\subsubsection{Controls} To provide more generalizability across ranges of stimuli, we include a suite of controls related to the projects (category, fundraising goal, description length, and quality) and participants (demographic characteristics, interest level in the projects, altruistic attitudes, and susceptibility to social influence).

\paragraph{Project category and description.} The projects we use are inspired by real crowdfunding efforts from the platform GoFundMe.com\footnote{\url{www.gofundme.com}}. Half of the chosen projects were successfully funded on GoFundMe, and half were not. We make sure that we have four different project categories (requests for homeless people, essential workers, students, and golf caddies) with associated funding goals between $\$5,000$ and $\$50,000$. The four project categories introduce critical variation in descriptions, covering a broad range of topics and target amounts. Within each project category with the same topic and target amount, we also pick four descriptions that are indistinguishable in terms of their main project-related factors that may influence whether people decide to contribute. Specifically, we ensure that the length of the description (as quantified by the number of words) and writing quality (as measured by the proselint\footnote{\url{www.proselint.com/lintscore}} score, Grammarly\footnote{\url{www.grammarly.com/}} (e.g., readability and total score), and Flesch–Kincaid readability tests~\cite{flesch2007flesch}) is similar. We also eliminate additional media content like images, provide the same details about project execution, use generic project names rather than real titles to make the projects unidentifiable, and remove location information to prevent regional bias. We thus select 16 descriptions in total. 

\paragraph{Participant demographics} We collect demographic data on participants' gender (male, female, non-binary, self-described), age (10-18, 19-29, 30-49, 50-64, $\geq$65), employment (student, seeking opportunities, employed part-time, employed full-time, retired, other), education (less than high school, high school, trade school, some college, bachelor's degree, advanced degree), income ($\leq$\$15k, \$15k--\$29k, \$30k--\$49k, \$50k--\$75k, \$75k--\$99k, $\geq$\$100k), and race (Asian, White, American Indian or Alaskan Native, Black or African American, and other). The demographics of the crowd workers closely mirror individuals who participate in crowdfunding, as reported on the crowdfunding platforms Kickstarter and Indiegogo\footnote{\url{https://artofthekickstart.com/crowdfunding-demographics-kickstarter-project-statistics/}}.

\paragraph{Interest level} We measure participants' baseline level of interest in the project category using a three-point Likert scale ``Not at all interested'' (0), ``Somewhat interested'' (1), ``Very interested'' (2). 

\paragraph{Altruism} To assess participants' altruistic tendencies, we use a self-reported altruistic personality scale~\cite{rushton1981altruistic}. This 20-item scale measures the frequency with which one engages in altruistic acts primarily toward strangers on a five-point scale ranging from ``Never'' (0) to ``Very Often'' (4). 

\paragraph{Susceptibility to social influence} We adapt scales from prior work investigating social influence in online social networks~\cite{stockli2020susceptibility}. Our adapted scale captures different facets of susceptibility to social influence (SSI), such as individuals' sensitivity to informative and normative influences~\cite{bearden1989measurement} and their tendency to seek information from others~\cite{reynolds1971mutually}. The 18-item SSI scale captures people's tendency to comply with social norms and to pay attention to others' behavior. For each item, we use a five-point Likert scale (``Strongly disagree'', ``Disagree'', ``Neither agree nor disagree'', ``Agree'', and ``Strongly agree'') as a continuous measure.

\subsection{Design of the Study}
\begin{figure}[!h]
    \centering
    \includegraphics[scale=.6]{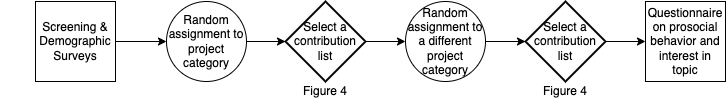}
    \caption{Outline of Study I: \emph{Single-Description Layout}. The study begins with a screening and demographics survey to ensure that participants are familiar with crowdfunding and are demographically representative of crowdfunders. Participants are then randomly assigned to two of four project categories and asked to select a contribution list from each. Participants conclude by completing a questionnaire on susceptibility to social influence, altruistic tendencies, and interest in the project category.}
    \label{fig:procedure_a}
\end{figure}

In Study I, we present participants with a mock crowdfunding setting that comprises one project description and two contribution lists that encode different crowd signals (high vs low variations in amounts and timing). We count the number of times that participants select each type of contribution list. After the participants select, we administer attention checks and ask them to reflect on their choice. Based on participants' selections, we perform statistical comparisons to determine whether variations in crowd signals influence people's likelihood to contribute to crowdfunding projects. To ensure robustness in our findings, we conduct the same analysis across different project categories and with different versions of the high and low variation in amounts and timing. We rely on qualitative coding of participants' open-ended responses to why they chose one contribution list over another to further understand why they are influenced by high or low variation in crowd signals. Finally, we perform one-way omnibus ANOVA tests accompanied by post-hoc Tukey tests to investigate whether participants' preferences for high or low variation are correlated with their susceptibility to social influence, altruism, or interest in a project category. Figure~\ref{fig:procedure_a} provides a sketch of the experimental procedure for this study. In what follows, we detail each step.

\subsubsection{Participants}
\begin{figure}[!h]
    \centering
    \includegraphics[scale=.4]{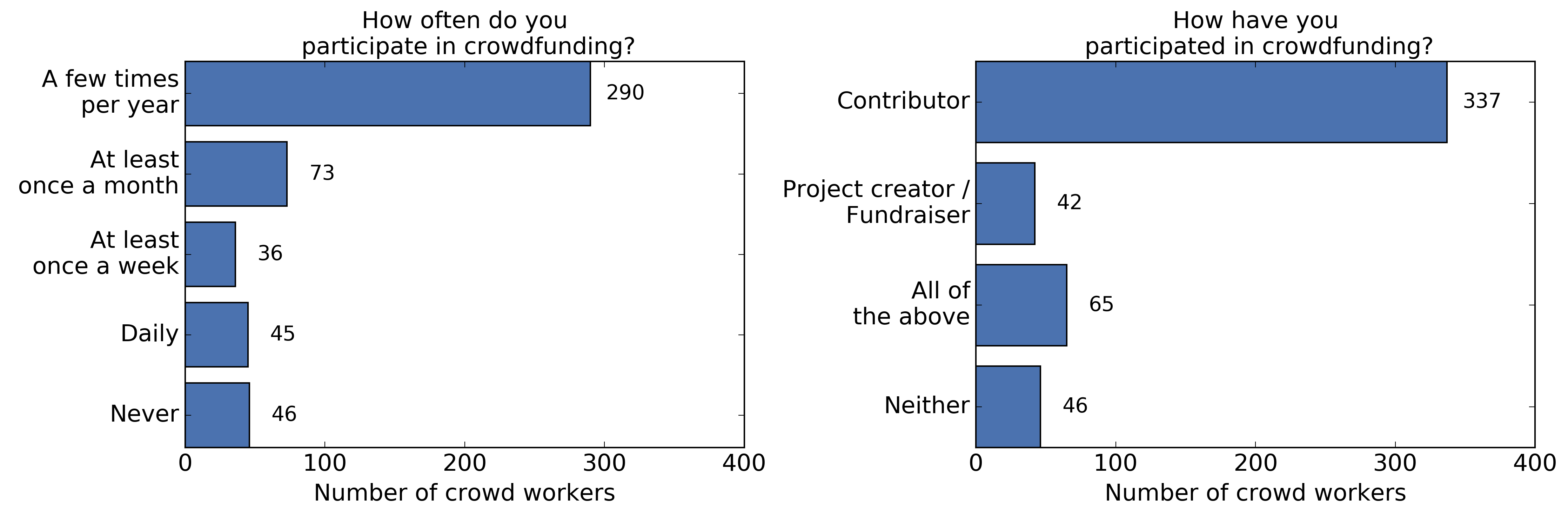}
    \caption{Participation in crowdfunding in Study I. Frequency of use of crowdfunding platforms (left). Type of use (right). Forty-six crowd workers (9.2\%) who indicate that they had never participated in crowdfunding and who did not use such platforms as a contributor or project creator were excluded from participating in the study.}
    \label{fig:familiarity_cf_exp1}
\end{figure}

We recruit English-speaking participants ($N=500$) through Amazon Mechanical Turk (MTurk). To ensure that the participants are representative of crowdfunders, we include screening questions at the beginning of the study (see Appendix) and \emph{only recruit crowd workers that} (i) demonstrate familiarity with the concept of crowdfunding, (ii) correctly identify examples of crowdfunding platforms, (iii) participate in crowdfunding at least a few times a year, and (iv) have contributed to \emph{or} created a crowdfunding campaign in the past. Figure~\ref{fig:familiarity_cf_exp1} shows a summary of the participants' frequency of participation and role in crowdfunding. Crowd workers who never participated in crowdfunding are excluded from the study after this screening survey. 454 participants (90.8\%) pass this screening on familiarity with crowdfunding and enroll in the study. Of these, 444 participants complete the study. Of the 444 participants who pass the crowdfunding familiarity screening and complete Study I, 60.8\% are male. Most of them are between 30 and 49 years old (66.4\%), Caucasian (73.6\%), have an annual income of less than $\$75,000$ (70.5\%), and hold a college or advanced degree (73.4\%). Participants are paid \$22.50 per hour upon completing the survey (including crowd workers that fail all attention check questions). The median task completion time is eight minutes for a \$3-pay per participant.

Considering that crowd workers typically come from specific socioeconomic backgrounds that may have implications for how they make decisions online~\cite{hargittai2020comparing,shaw2021online}, we ensure that the demographics of the crowd workers in our sample of participants closely mirror individuals who take part in crowdfunding~\cite{morejon2016crowdfunding}. 
To assure data reliability, from the beginning, we restrict the survey to participants with a Human Intelligence Task (HIT) approval rate greater than 98\% to be consistent with research relying on MTurk~\cite{berinsky2012evaluating,goodman2013data,paolacci2010running,peer2014reputation}. We include filters to confirm that participants had not taken prior crowdfunding surveys from our team, are above 18, and are capable of consent. Finally, to ensure that participants observe the same study design layout, we only admit those using a desktop or laptop computer, not a mobile device.

\subsubsection{Randomization and Counterbalancing}
After completing the screening and demographic survey, participants are randomly assigned to two of four project categories. For each project category, we prepare four different project descriptions and two versions of crowd signals for high vs low variation in contribution amounts and contribution timing. As a result, we create 16 different combinations (4 x 2 x 2) for use in the selection task for each project category. Participants perform the selection task twice. In each task, they see project listings from a different project category.

\subsubsection{Pretesting Mock Crowdfunding Project Descriptions} 
After selecting the four project descriptions within each of the four categories (homeless people, goal: \$5,000; golf caddies, goal: \$50,000; PPE for essential workers, goal: \$10,000; high school robotics competition, goal: \$5,000), we pretest them with $N=136$ MTurkers. The pretest reveals that people are highly sensitive to the slightest differences in language. For instance, they pick up on certain words that they prefer over others, even though the requests are similar. Therefore, we manually edit the descriptions to make them as similar as possible within each category until people's preferences for different descriptions are uniform.

\subsubsection{Contribution Lists as Realistic Signaling Scenarios}
\begin{figure}[!h]
    \centering
    \includegraphics[scale=.5]{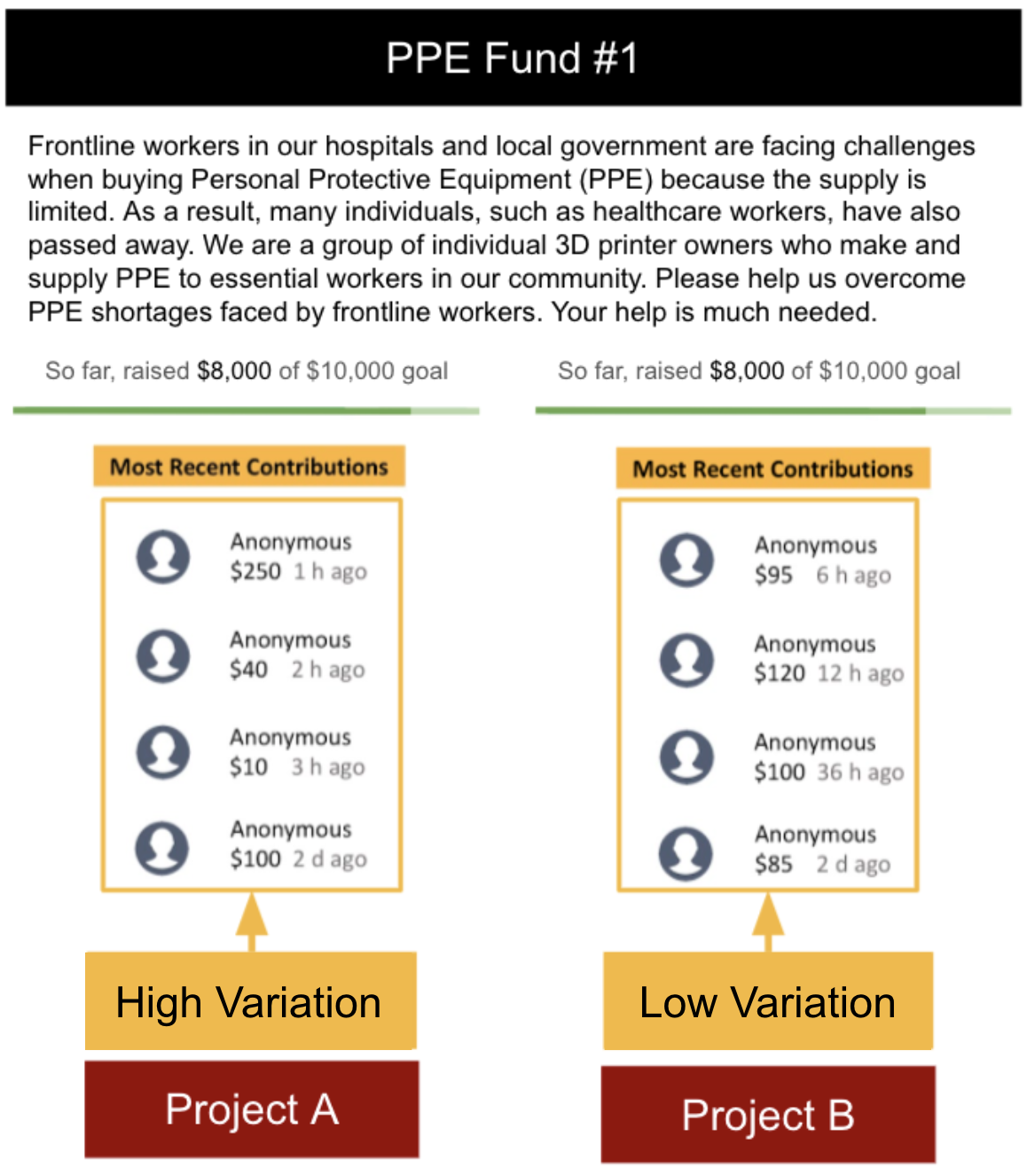}
    \caption{Example task presented to participants in Study I. Each participant sees a project description with two different contribution lists with high variation, or heterogeneity, in contribution amounts and time and with low variation, or homogeneity, in contribution amounts and time. Project category in this example: Supporting the manufacture of Personal Protective Equipment (PPE) with 3D-printing.}
    \label{fig:exp_1}
\end{figure}

Crowdfunding platforms contain contribution lists that show users when and how much others have already contributed to a specific project. When creating artificial contribution lists, we focus exclusively on the crowd signals in our two treatment conditions (high vs low variation in contribution amount and contribution timing). Consistent with the formatting of most crowdfunding platforms that display a few contributions for each project, every list contains the same number of most recent contributors (four). Additionally, we ensure that the four recent contribution amounts add up to \$400 in each contribution list so that every list appears to be raising the same amount of money. We choose to show that 80\% of the target amount has been raised to indicate that all projects have an equal chance of success. This way, our contribution lists encode non-straightforward characteristics of prior contributions that reflect high and low variation in amounts and timing but are otherwise indistinguishable (Figure~\ref{fig:exp_1}).

\subsubsection{Reflection Questions} To explore why people choose a project in one treatment condition over another, we ask participants to provide their reasoning \emph{after} the selection. We study the responses using thematic analysis~\cite{braun2006using}. This method entails a bottom-up, inductive approach starting with open coding of all the data into themes. Some of the emerging themes are related to the treatment conditions of high or low variation; others are not. The resulting five themes encompass participants' reasons to contribute based on subtle differences in the project \emph{descriptions}, different crowd treatment (i.e., \emph{high variation} and \emph{low variation}), selecting projects at \emph{random}, and for \emph{other} reasons that were difficult to group into a meaningful category.

We develop a codebook to enable multiple annotators to label participants’ open-ended responses into the above categories. Two independent annotators code 20\% of the data and discuss disagreements. After this discussion, the annotators code a different 35\% of the corpus and agree with Cohen's $k=0.71$. Since differences between reasons for choosing the heterogeneous vs homogeneous contributions and times are crucial to our study, we conduct a more in-depth thematic analysis of the responses grouped in these two themes. We find responses that align with the conditional cooperation theory, the underdog effect, and the theory of impact philanthropy as introduced in Section~\ref{related-work}.

\subsubsection{Post-Survey Questionnaire}
To investigate whether contribution lists corresponding to high vs low variation in contribution amounts and timing could encode other factors that might influence fundraising outcomes~\cite{steigenberger2017supporters,rodriguez2019altruism,giudici2018reward}, we include a post-survey questionnaire to evaluate the effect of crowd signals depending on participants' level of interest in a project category, their altruistic tendencies, and susceptibility to social influence. 

\subsubsection{Attention Checks}
We incorporate attention checks into the survey to ensure participants pay attention to the visual information displayed in the contribution lists. They complete the attention check questions \emph{after} finalizing the selection task when they can no longer see the project selection screen. To pass the attention check questions, participants must recall how many contributions are visible on the project description page, the project's fundraising goal, and how much money the campaign has raised when this information is no longer available to them. These demanding attention checks are more stringent than most commonly used instructional manipulation checks (IMC) like ``you should not answer this question if you read it; it is to check your attention.'' While the chosen attention checks result in a much higher participant task failure rate (15.7\%) compared to other MTurk studies that employ IMCs, the reported failure rate is comparable with survey studies. It ensures that participants indeed perceive all the aspects of the information that characterize our treatment conditions~\cite{paas2018instructional,hauser2016attentive,oppenheimer2009instructional}. We consider task responses from participants that pass at least one attention check on each task.

\subsection{Results}
\subsubsection{More Participants Choose Contribution Lists with High than Low Variation in Amounts and Timing} We investigate whether observing high crowd signals makes people more likely to select a contribution list. We count the cases when participants select high variation in crowd signals (329, corresponding to 56.5\%) vs low variation (253, corresponding to 43.5\%). Thus, a project is 29.89\% more likely to attract contributors when assigned a contribution list with high rather than low variation in the amounts and timing of contributions. This implies that participants systematically choose projects with high variation over those with low variation ($\chi^2$(1, $N=582$) = 9.924, $p = 0.0016$).

\begin{table}[!h]
\caption{Participants' selections by project category and fundraising goal in Study I. More respondents preferred projects with high than low variation in all four project categories.}
\centering
\begin{tabular}{|l|c|c|}
\hline
Category & Treatment & Number (\%) of Selections \\ \hline
\multicolumn{1}{|l|}{Caddies} &         High Variation & 78 (53.4\%) \\
\multicolumn{1}{|l|}{Goal: \$50,000} & Low Variation & 68 (46.6\%) \\ \hline
\multicolumn{1}{|l|}{Homelessness} &    High Variation & 83 (56.8\%)  \\
\multicolumn{1}{|l|}{Goal: \$5,000} &  Low Variation & 63 (43.2\%) \\ \hline
\multicolumn{1}{|l|}{PPE} &             High Variation & 91 (61.5\%)  \\
\multicolumn{1}{|l|}{Goal: \$10,000} & Low Variation & 57 (38.5\%)  \\ \hline
\multicolumn{1}{|l|}{STEM} &            High Variation & 77 (54.2\%)  \\
\multicolumn{1}{|l|}{Goal: \$5,000} &  Low Variation & 65 (45.8\%)  \\ \hline
% \multicolumn{1}{|l|}{Total} & High \&   Low & 582 & 615 & 1,197 \\ \hline
\end{tabular}
\label{tab:sel-category-1}
\end{table}

\paragraph{The preference for high variation in crowd signals is stable across different project categories and target amounts} To investigate whether participants' preference for high crowd signals is influenced by differences in project categories or fundraising goals, we analyze selections separately in each project category. Across different categories with fundraising goals ranging from \$5,000 to \$50,000, participants consistently choose projects with high variation in contribution timing and amounts (Table~\ref{tab:sel-category-1}). They prefer the high variation over low variation condition 14.7\% more often in the golf caddies category, 31.7\% in homelessness, 59.6\% in PPE, and 18.5\% in STEM. The differences between categories are substantial and do not seem to correlate with the target amount. The preference for high variance in crowd signals is thus robust to different crowdfunding project categories and fundraising goals.

\paragraph{Participants prefer the high over low variation condition in open-ended responses} When further asked why they selected one contribution list over another, most participants (66.4\%) mention that they chose based on the crowd signals. For example, ``\textit{I liked the wider range of contribution amounts in the second list}'' or ``\textit{I choose the Robotics Team Travel Fund \#3 because, their contribution list was a bit less consistent in terms of donation amounts and time frame.}.'' Participants produced these explanations \emph{without any prompts} that would have guided them to provide answers related to the consistency or variation in amounts and times. 

\begin{figure}[!h]
    \centering
    \includegraphics[scale=.35]{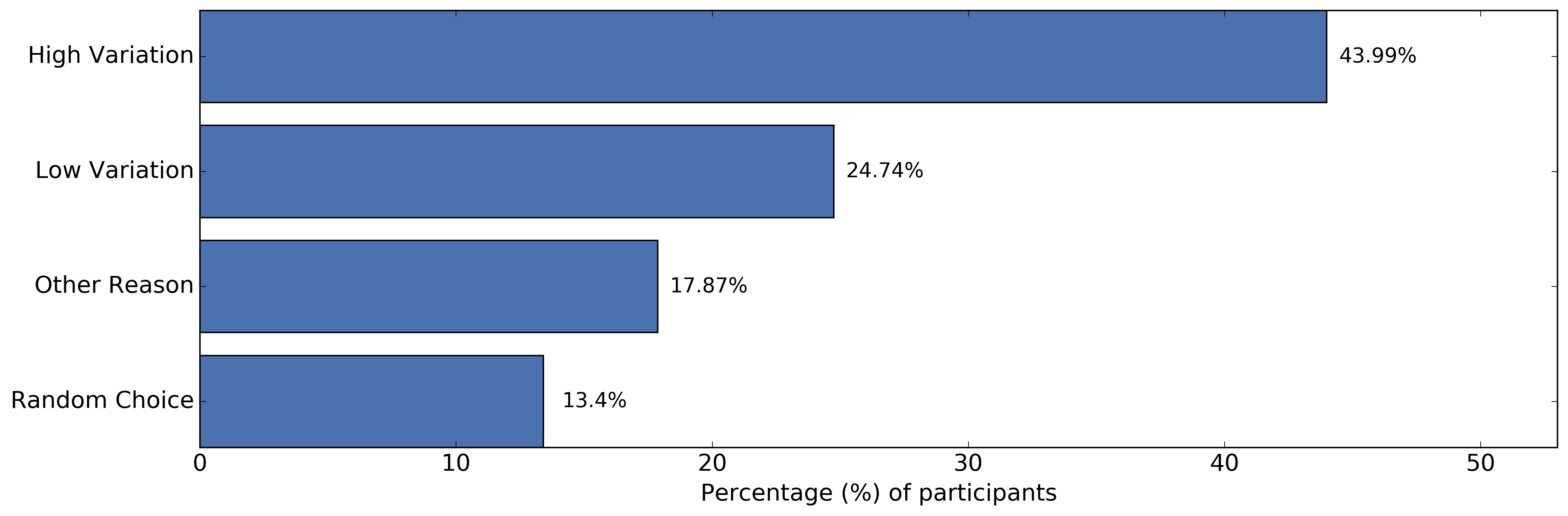}
    \caption{Percentage of participants (x-axis) that made their selections based on the reasons indicated on the y-axis. The research team coded the reasons using thematic analysis. Most participants attribute their decision to variation in crowd signals, and systematically prefer the condition with high variation in contribution amounts and timing.}
    \label{fig:reasons_a}
\end{figure}

Importantly, according to their free text responses, more participants prefer the high variation treatment condition (43.99\%) than the low variation treatment condition (24.74\%). A smaller group of participants (13.4\%) indicated that they selected a list at random (e.g., ``\textit{It was a toss-up--they both were equally funded with pretty similar amounts}'') and the remaining 17.1\% provided reasons that could not be classified clearly into any of the above categories (e.g., ``\textit{I preferred the first one}''). Figure~\ref{fig:reasons_a} provides a summary of these findings. Next, we unpack the specific reasons given by participants for choosing the treatment condition with high (Section~\ref{sec:whyhigh}) and low variation (Section~\ref{sec:whylow}).

\subsubsection{Reasons Why Participants Choose High Variation in Crowd Signals} \label{sec:whyhigh}
High variation crowd signals imply contributions of varying amounts provided in unequally spaced time intervals. Participants' reasons for selecting such contribution lists include (1) perceptions of broad project appeal, (2) feeling comfortable to give at any contribution level, and (3) the underdog effect.

\paragraph{Perceptions of broad project appeal} According to some participants, the presence of both small and large contribution amounts (e.g., \$10 and \$250) makes a fundraising campaign feel more like a grassroots effort. Involving a diverse community of funders has been shown via interviews to be essential for attracting potential contributors~\cite{hui2014understanding}. Our study also reveals that the presence of varied contribution amounts make the participants feel like there is more freedom in the amount that one could contribute, which signals a project's broad appeal across a wide range of contributors. For example, one participant noted:
\begin{quote}
    -- Broad Appeal Example: \textit{``The list I chose had donation amounts that varied from 10 dollars to 250 dollars. This makes me feel like there is more freedom in the amount you are expected to donate and makes me more willing to donate.''}
\end{quote}

\paragraph{Feeling comfortable to give at any contribution level} The presence of small contribution amounts in the high treatment condition makes some participants feel more comfortable giving any amount. Participants noted that small contributions made a campaign more inviting to contributors. This explanation is particularly common among participants who would only be willing to contribute a small amount to the project and do not want their contribution to stand out in the low variation list with consistently high contributions (e.g., in the \$80 -- \$120 range). Additionally, by contributing a small amount to the high variation contribution list, some participants felt that their contribution would not be seen as too little but rather valued:

\begin{quote}
    -- Feeling Comfortable Example \#1: \textit{``I went with the first one because the donations were all over the place and there was (sic) some low ball donations and my donation wouldn't look as weak if I went on that list.''}
\end{quote}

\begin{quote}
    -- Feeling Comfortable Example \#2: \textit{``Because it has smaller amounts listed and I don't want my smaller contribution listed around bigger dollar amounts because even if it is anonymous I don't want to do that. It feels embarrassing.''}
\end{quote}

\paragraph{The underdog effect} Compared to the low treatment condition which includes consistently high contribution amounts, some participants interpreted the presence of small amounts in the high treatment condition as a signal that a project was struggling to raise funds and thus needed more help. Consistent with research about how funders are willing to contribute towards underdog (vs top dog or neutral) narratives or to support charity projects with low prior support compared to well-funded projects~\cite{zamudio2018e13,bradley2019relatively,duncan2004theory}, some participants in our study showed a higher level of empathy and altruism towards the high treatment condition that included low contribution amounts:

\begin{quote}
    -- Underdog Effect Example \#1: \textit{``The right side has less money from its contributors. So I would like to contribute more to help the project going.''}
\end{quote}

\begin{quote}
    -- Underdog Effect Example \#2: \textit{``I chose one list over the other because the most donations made were small and seemed like they needed more donations.''}
\end{quote}

\subsubsection{Reasons Why Participants Choose Low Variation in Crowd Signals} \label{sec:whylow}
Low variation in crowd signals implies roughly equal contributions at consistent time intervals. Participants' reasons for selecting such contribution lists include (1) conditional cooperation or a willingness to contribute conditional on others' contributions, (2) consensus on a project’s merit among funders, (3) perceptions of the predictability of the project's outcome, and (4) projection of generosity.

\paragraph{Conditional cooperation} Conditional cooperation occurs in public goods settings when people tie their contribution to others' support~\cite{fischbacher2001people}. Accordingly, we expect crowdfunding participants to prefer the low variation condition that showcases consistent amounts. Such amounts suggest an equal sharing of the burden, which might satisfy participants' fairness preferences, as highlighted by the following two responses:

\begin{quote}
    -- Conditional Cooperation Example \#1: \textit{``I saw a good spread around the same dollar amount contributed. This shows me all the people are equally passionate, this matters a lot to me.''}
\end{quote}

\begin{quote}
    -- Conditional Cooperation Example \#2: \textit{``I felt like the amounts in the 2nd list were more even and steady compared to the 1st list. The 1st list gave me concern with over half of [the] total in the recent 4 donations came from 1 single donation. I would feel better adding my donation to that second list.''}
\end{quote}

\paragraph{Consensus on a project's merit} Some participants pointed out the similarities in the contribution amounts as their main reason for choosing the low variation list. These participants interpret the presence of uniform amounts and timing as a signal of consistency, confidence, or consensus among the funders about the projects' merit:

\begin{quote}
    -- Consensus Example \#1: \textit{``The donation amounts were all of a similar amount, so it seemed like there was a consensus on the project.''}
\end{quote}

\begin{quote}
    -- Consensus Example \#2: \textit{``I thought list A showed a more even investment amongst all members than B did, meaning more people thought it was an idea worth throwing a significant amount of money at.''}
\end{quote}

\paragraph{Perceived predictability of project outcome} Consistently high contributions also make a project seem more likely to succeed, given that the recent contributors are willing to pitch in large amounts and hence must have believed the project was worth investing significant funds into. Therefore, some participants perceive the consistently high amounts in the low variation list to be more promising than the smaller amounts in the high variation list:

\begin{quote}
    -- Predictability Example \#1: \textit{``More `equal' contributions across the list feels like the project has a very predictable merit across a range of people. The other list felt very random and not as compelling.''}
\end{quote}

\begin{quote}
    -- Predictability Example \#2: \textit{``I chose the first one because it has figures that remain stable and have a better chance of continuing to contribute a high fixed amount in the future.''}
\end{quote}

\paragraph{Projection of generosity} Finally, some participants that choose the low variation condition express the desire to do their part in a way that projects generosity. Such giving contributes to people's sense of warm-glow, which refers to the emotional reward of giving to others~\cite{xu2022implications}. We observe that several participants who choose the low variation condition appreciate others' consistent contributions and are compelled to give at similarly high levels of generosity. As two participants explained: 

\begin{quote}
    -- Generosity Example \#1: \textit{``I chose the first one because the average overall seemed high. I would like to be a part of that one to show I can give a high amount, as well.''}
\end{quote}

\begin{quote}
    -- Generosity Example \#2: \textit{``I chose the second one because the overall average is high. I wanted to contribute a similar high amount so I can make the overall contribution look good.''}
\end{quote}

\subsubsection{The Influence of Crowd Signals Is Most Substantial among Participants Susceptible to Social Influence} To better understand whether participants' motivations and predispositions are associated with their selections in Study I, we group subjects based on their reasons for choosing one contribution list over the other. Conducting a one-way omnibus analysis of variance (ANOVA), we observe a significant difference in the distribution of social influence scores ($F=29.731, p<0.001$), altruism scores ($F=8.246, p<0.001$), and baseline interest in a project category ($F=5.845, p=0.001$) across the groups of high variation, low variation, and random choice. To further investigate which of the three groups is significantly different from the others, we perform multiple post-hoc pairwise comparisons using Tukey's Honest Significant Difference (HSD) test. 

Across groups of people who prefer high or low variation in crowd signals, we observe no statistically significant differences in the averages of their social influence scores, altruism scores, or baseline interest in a project category. Hence participants' responses to heterogeneous vs homogeneous contribution amounts and timing are not affected by susceptibility to social influence, altruism, or interest in the project topic. However, participants who select projects at random (e.g., ``\textit{I chose randomly as I didn't see which choice was different than the other.}'') are the least susceptible to social influence and show significant differences in the averages of their social influence scores in comparison with participants that are influenced by the crowd signals, either low or high (Table~\ref{tab:tukey_a}). We observe no significant difference in participants' altruism scores or baseline interest in a project topic in pair-wise comparisons. 

\begin{table}[!h]
\caption{Study I: Post-hoc Tukey Honestly Significant Difference (HSD) test results for one-way ANOVA on participants' susceptibility to social influence across groups who state in the open-ended questions that their reason for choosing the contribution list was related to heterogeneous contribution amounts and timing, i.e., high variation in crowd signals (``High''), homogeneous amounts and timing or low variation (``Low''), or was random (``Random''). P-values significant at: p<0.001***, p<0.01**, and p<0.05*.}
\label{tab:tukey_a}
\centering
\begin{tabular}{r|llllll|}
\hline
\multicolumn{1}{|l|}{\textbf{Susceptibility to Social Influence}} & \multicolumn{1}{c|}{\begin{tabular}[c]{@{}c@{}}mean\\ (group1)\end{tabular}} & \multicolumn{1}{c|}{\begin{tabular}[c]{@{}c@{}}mean\\ (group2)\end{tabular}} & \multicolumn{1}{c|}{diff} & \multicolumn{1}{c|}{se} & \multicolumn{1}{c|}{T} & \multicolumn{1}{c|}{p-tukey} \\ \hline
\multicolumn{1}{|l|}{High - Low} & \multicolumn{1}{l|}{31.434} & \multicolumn{1}{l|}{34.264} & \multicolumn{1}{l|}{-2.830} & \multicolumn{1}{l|}{1.439} & \multicolumn{1}{l|}{-1.967} & 0.201 \\ 
\multicolumn{1}{|l|}{High - Random *} & \multicolumn{1}{l|}{31.434} & \multicolumn{1}{l|}{26.372} & \multicolumn{1}{l|}{5.062} & \multicolumn{1}{l|}{1.787} & \multicolumn{1}{l|}{2.833} & 0.024 \\
\multicolumn{1}{|l|}{Low - Random ***} & \multicolumn{1}{l|}{34.264} & \multicolumn{1}{l|}{26.372} & \multicolumn{1}{l|}{7.892} & \multicolumn{1}{l|}{1.942} & \multicolumn{1}{l|}{4.064} & 0.001 \\ 
\hline
\end{tabular}
\end{table}

\subsection{Summary of Study I and Outlook to Study II}
Study I tests the fundamental signaling mechanism present in crowdfunding. It shows that high variance in the amounts and timing of contributions attracts significantly more funders than low variance \emph{when} participants choose from two contribution lists for the \emph{same} project description. We demonstrate that signaling impacts more those who are susceptible to social influence. Finally, the role of signaling in decision-making does not depend on the project category and topic, fundraising goal, the written request's length and quality, participants' interest level in the projects, or participants' altruistic attitudes. These results indicate that the signaling mechanism is robust across various stimuli and conditions. 

In the course of designing and running Study I, we make three observations. First, through the direct comparison of contribution lists, participants are primed to observe some role of crowd signals. Second, the decision setting is relatively unrealistic because people typically observe multiple projects on crowdfunding platforms. Third, Study I does not directly control for the project description, although this has a significant effect on people's decision-making~\cite{althoff2014ask,mitra2014language,larrimore2011peer,rhue_emotional_2018,anglin2018power}. These observations propel us to investigate the signaling mechanism in a more realistic context. With Study II, we probe the efficacy of the signaling mechanism when different signals are attached to different project descriptions, providing a more stringent test for the effect of crowd signals.

\section{Study II: Mechanisms in Context: The Layout with Two Descriptions} 
\begin{figure}[!h]
    \centering
    \includegraphics[scale=.6]{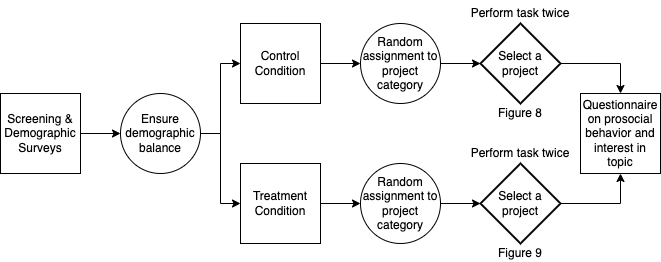}
    \caption{Outline of Study II: \emph{Layout with Two Descriptions}. The study begins with a screening and demographic survey to ensure that participants are familiar with crowdfunding and are demographically representative of crowdfunders. Participants are split between control and treatment groups, randomly assigned to one of four project categories, and perform a selection task twice. Participants conclude by completing a questionnaire on susceptibility to social influence, altruistic tendencies, and interest in the project category.}
    \label{fig:procedure_b}
\end{figure}

To investigate whether the results of Study I are consistent when we use a layout that emulates the presence of competing projects, we employ a setup that enables us to directly compare two signaling scenarios, each with its own project description. Our randomization assures that the same descriptions are shown to participants both with heterogeneous and homogeneous contribution amounts and timing, allowing us to compare how frequently a description is chosen depending on whether it appears in the high vs low variation condition. 
In this study, we split participants into treatment and control groups, as shown in the sketch in Figure~\ref{fig:procedure_b}. We use the same measures and analysis methods as in Study I (see Section~\ref{sec: measures}).

\subsection{Design of the Study}
\subsubsection{Participants}
\begin{figure}[!h]
    \centering
    \includegraphics[scale=.4]{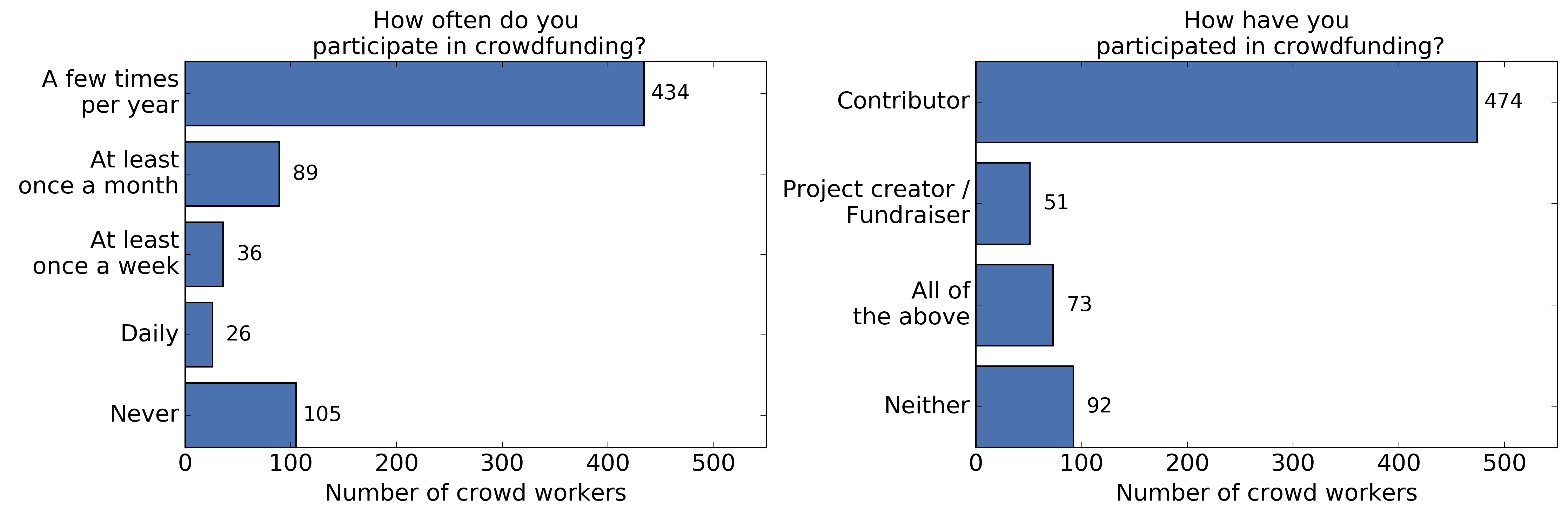}
    \caption{Participation in crowdfunding in our sample of MTurkers in Study II. Frequency of use of crowdfunding platforms (left). Type of use (right). One hundred fourteen crowd workers (15.2\%) who indicated that they have never participated in crowdfunding and who did not use such platforms as a contributor or project creator were excluded from participating.}
    \label{fig:familiarity_cf_exp2}
\end{figure}

For Study II, we recruit $N=750$ participants, $250$ for the control group and $500$ for the treatment condition. We include the same screening questions used in Study I at the beginning of the study (see Appendix) to ensure that participants are representative of crowdfunders. Figure~\ref{fig:familiarity_cf_exp2} shows the corresponding summary of the participants' responses to questions about their role and frequency of participation in crowdfunding in Study II. Of the $750$ crowd workers that attempt the screener, 636 participants (84.8\%) pass the screening on familiarity with crowdfunding and are enrolled in the study as participants. Of the 636 participants enrolled, 576 participants complete the survey. Three hundred and eighty-six participants (67\%) are in the treatment group, and 190 participants (33\%) are in the control group. Of all the admitted participants in the treatment and control groups, 51.4\% are male. Most of them are between 30 and 49 years old (63.9\%), Caucasian (76.0\%), have an annual income less than $\$75,000$ (66.8\%), and have a college or advanced degree (73.8\%). They are compensated equally to the participants in Study I.

\subsubsection{Study Procedure}
As shown in Figure~\ref{fig:procedure_b}, participants are randomly assigned to either the control or treatment condition after completing the screening and demographic survey. To maintain the demographic balance between the treatment and control groups, we ensure that each group has an equal number of participants across different demographic categories such as gender, age, income, education, etc. Participants are further randomly assigned to one of four project categories. Similar to Study I, for each project category, we have 16 combinations (four project descriptions times four versions of contribution amounts and timing, as shown in Table~\ref{tab:artficial-signals}). During each of the two selection tasks, participants are presented with two, non-overlapping project listings and asked to choose which one they are more likely to contribute to. The setup ensures thus that each participant reads all four project listing descriptions that are prepared for a given project category by the end of the two selection tasks.

\begin{figure}[!h]
    \centering
    \includegraphics[scale=.75]{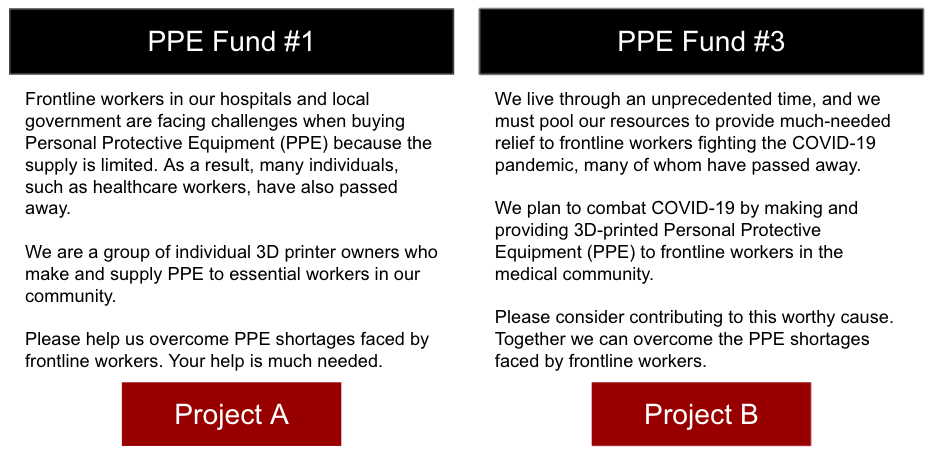}
    \caption{Example task presented to participants in the \emph{control} group of Study II. Each participant sees two project descriptions without any contribution lists. Project descriptions shown in this example are from the PPE category.}
    \label{fig:exp_2_control}
\end{figure}

In the control condition, each selection task presents two similar project descriptions side-by-side without contribution lists, i.e., there are no contribution lists and hence there is no indication of crowd signals (see Figure~\ref{fig:exp_2_control}). Much like when we pre-test the descriptions, this setup allows us to collect the baseline appeal of all campaigns among our participants. 

\begin{figure}[!h]
    \centering
    \includegraphics[scale=.325]{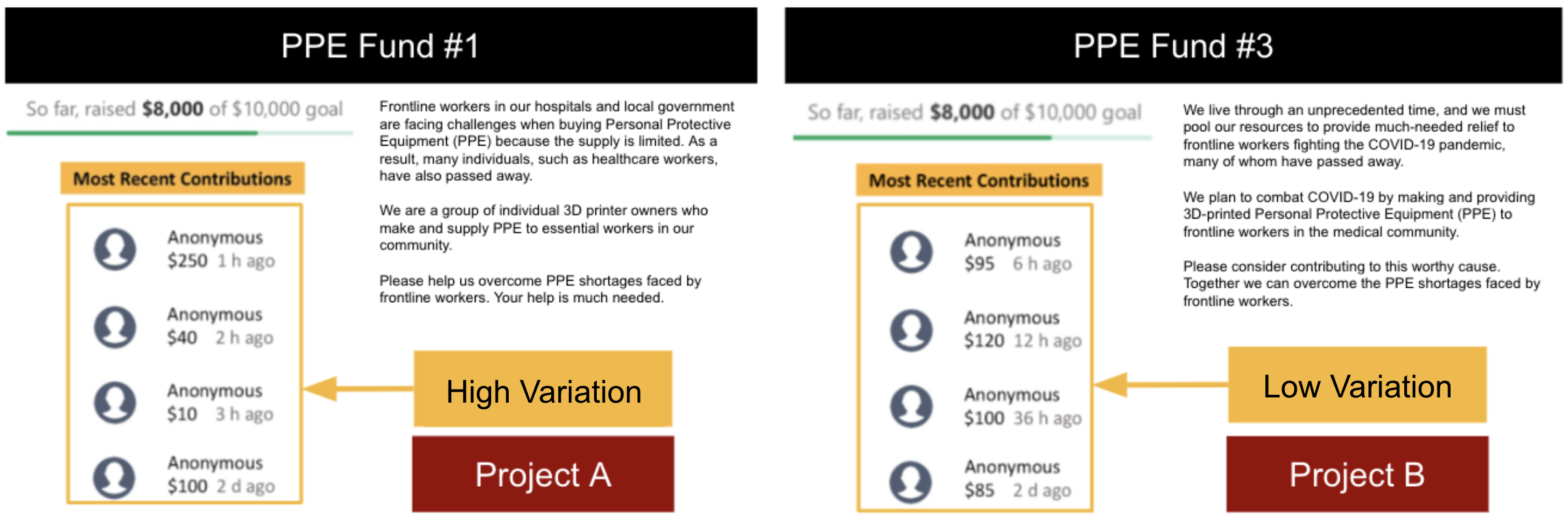}
    \caption{Example task presented to participants in the \emph{treatment} group of Study II. Each participant sees two similar project descriptions with two different contribution lists, one displaying high variance in contribution amounts and timing and one representing low variance in amounts and timing.}
    \label{fig:exp_2_treatment}
\end{figure}

In the treatment condition, we provide information about crowd features using the same artificial contribution amounts and times as in Study I (Table \ref{tab:artficial-signals}). However, each selection task presents two similar project descriptions from the same project category side-by-side. One of the descriptions is followed by a contribution list with low variation in crowd signals (homogeneous contribution amounts and timing), and the other is presented with a contribution list with high variation in crowd signals, i.e., heterogeneous contribution amounts and timing (see Figure~\ref{fig:exp_2_treatment}).

We perform attention checks and ask participants to reflect on their choice. Using the same codebook as in Study I, two independent annotators code 37\% of the corpus and reach an agreement of Cohen's $k=0.80$.

Finally, participants in both the control and treatment groups fill out our questionnaire about their motivations and predispositions to altruism and social influence.

\subsection{Results} \label{sec:exp2results}

\subsubsection{Participants Have Similar Preferences for Different Project Descriptions} To test for possible confounding effects introduced by the project descriptions, we count how often each project description is selected in our \textit{control} condition. We find that the projects' baseline appeal ranges from 30.6\% to 65.3\%, with most of the descriptions likely chosen at random (average = 50.0\%, standard deviation = 10.0). Additionally, we find no significant relationship between the projects' baseline appeal in the control condition and participants' preferences in the treatment condition. This finding confirms that the competing project descriptions do not confound the influence of the crowd signals.

\subsubsection{More Respondents Choose Contribution Lists with High than Low Variation in Contribution Amounts and Timing} Participants in the treatment condition of Study II select high variation in amounts and timings in 322 (52.4\%) cases versus low variation in 294 (47.6\%) cases. This means that, in the context of competing topic-oriented information, a project is 10.1\% more likely to attract funders when assigned a contribution list with high rather than low variation ($\chi^2$(1, $N=582$) = 1.272, $p = 0.259$). This trend is consistent across different project categories, except for the golf caddy project category, which aims to raise \$50,000 (Table~\ref{tab:sel-category-2}). We believe this unique result might be due to a lower interest in the project category (average=1.565, standard deviation=0.627) compared to the other three categories (average=2.398, standard deviation=0.632). For example, when asked why they selected one golf caddies project over another, one participant noted, ``\textit{I don't really care about caddies so I didn't think that much.}'' This decreased interest might result in participants not paying close attention to the selection task in this specific category.

\begin{table}[!h]
\caption{Participants' selections by project category and fundraising goal in Study II. More respondents preferred projects with high than low variation in contribution amounts and timing in all project categories, but golf caddies.}
\centering
\begin{tabular}{|l|c|c|}
\hline
Category & Treatment & Number (\%) of Selections \\ \hline
\multicolumn{1}{|l|}{Caddies} &         High Variation &  62 (44.9\%)  \\
\multicolumn{1}{|l|}{Goal: \$50,000} & Low Variation &  76 (55.1\%)  \\ \hline
\multicolumn{1}{|l|}{Homelessness} &    High Variation &  102 (55.1\%)  \\
\multicolumn{1}{|l|}{Goal: \$5,000} &  Low Variation &  83 (44.9\%)  \\ \hline
\multicolumn{1}{|l|}{PPE} &             High Variation &  56 (50.9\%)  \\
\multicolumn{1}{|l|}{Goal: \$10,000} & Low Variation &  54 (49.1\%)  \\ \hline
\multicolumn{1}{|l|}{STEM} &            High Variation &  102 (56.0\%)  \\
\multicolumn{1}{|l|}{Goal: \$5,000} &  Low Variation &  80 (44.0\%) \\ \hline
% \multicolumn{1}{|l|}{Total} & High \&   Low & 582 & 615 & 1,197 \\ \hline
\end{tabular}
\label{tab:sel-category-2}
\end{table}

\subsubsection{Observing Two Project Descriptions Slightly Decreases the Effect of Crowd Signals} We investigate the extent to which observing two similar project descriptions affects the influence of having heterogeneous contribution amounts and timing (high variation) as opposed to homogeneous contribution amounts and timing (low variation). To this end, we perform a Chi-squared test that determines whether distributions of participants' project selections by high vs low variation in crowd signals differ between Studies I and II. Our findings show that observing simultaneously two descriptions instead of one slightly lowered the percentage of participants' selecting the high variation condition in Study II (52.4\%) compared to Study I (56.5\%) across all project categories ($\chi^2(1,N=1197) = 11.97, p<0.01$). Therefore, basic competition between projects has a small effect ($\phi=0.099$) on crowd signals\footnote{Phi $\phi$ is a measure of effect size for Chi-square tests $\chi^2$ and is defined as $\frac{\chi^2}{n}$. A value of 0.1 is considered a small effect, 0.3 a medium effect, and 0.5 a large effect.}.

\subsubsection{Most Participants Attribute Their Choice to Project Descriptions, which Are Virtually Indistinguishable} In contrast to Study I, where participants select one of two contribution lists belonging to the same project description, participants in Study II select one of two separate project descriptions followed by a high and low variation contribution list. Participants in Study II, therefore, evaluate both the information presented in the project descriptions \emph{and} the crowd signals encoded in the contribution lists. When asked for the reasons behind their selections, most of them (54.83\%) mention deciding based on project descriptions. Specifically, they attribute their choice to differences between descriptions. For example, a couple of participants noted:

\begin{quote}
    -- Description Example \#1: \textit{``The wording on the second one was much clearer and more direct about the issues facing the homeless population. I also appreciated that it said `if you feel compelled to help,' making it seem like the choice was up to me.''}
\end{quote}

\begin{quote}
    -- Description Example \#2: \textit{``The write up had a more natural and friendly tone to it. It felt like [project A] was aimed directly at me and felt like an actual conversation with someone.''}
\end{quote}

This is very interesting because the differences between the project descriptions are negligible. We edit requests to ensure that they are similar in length, writing quality, and readability scores (see Section~\ref{sec: measures}). The results on the control condition confirm that we successfully made the descriptions uniform. Additionally, even the slightest differences would be negligible due the randomization process. Given this randomization, participants could not have chosen based on the linguistic features of the descriptions. Indeed, the same project description is described as ``more friendly/natural'' by participants who select the high variation in contribution amounts and times, and ``less friendly/natural'' by those who select the low variation signals. 

Critically, \emph{participants' own reflections suggest that most of them are unconcerned with any encoded crowd signals, yet they still systematically choose contribution lists with heterogeneous amounts and timing}. This finding is crucial, given that participants are shown counterbalanced layouts in crowd signals and project-related controls. 

\begin{figure}[!h]
    \centering
    \includegraphics[scale=.35]{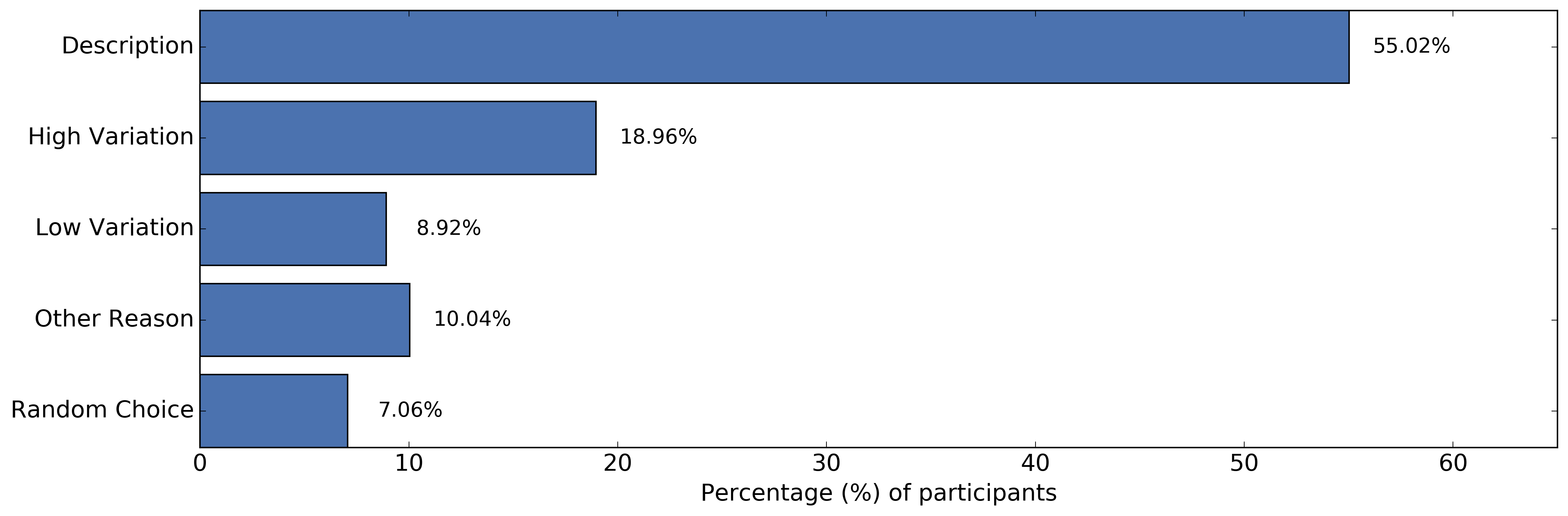}
    \caption{Percentage of participants (x-axis) that made their selections based on the reasons indicated on the y-axis. The reasons have been coded by the research team using thematic analysis. Most participants attribute their decision to the description even as they systematically choose the condition with high variation in crowd signals.}
    \label{fig:reasons_b}
\end{figure}

We find that across the various counterbalanced designs, 27.88\% of the participants mention the crowd signals as their decision guide. Consistent with Study I, more participants state that they prefer the high variation condition (18.96\% choose heterogeneous contribution amounts and timing) over the low variation condition (8.92\% choose homogeneous contribution amounts and timing). A small group of participants (7.06\%) choose at random between the two projects, and the remaining 10.04\% provide other reasons for their decision that could not be classified into the above categories (Figure~\ref{fig:reasons_b}).

\subsubsection{Participants Who Are Responsive to Project Description and Crowd Signals Are More Susceptible to Social Influence than Those Making Random Selections} As in Study I, we test for relationships between participant choices and measurements of susceptibility to social influence, altruism, and interest in the project category. Consistent with the first study, an omnibus one-way ANOVA shows a significant difference in the social influence scores ($F=7.217, p<0.001$) of those who self-report making decisions based on project descriptions, crowd signals, and randomly. 

To investigate which groups are significantly different from each other, we perform multiple post-hoc pair-wise comparisons using Tukey’s Honest Significant Difference (HSD) test (Table~\ref{tab:tukey_b}). We observe significant differences in susceptibility to social influence (SSI) scores between participants that select at random and those who choose based on either the description ($t = 3.443, p=0.005$), or high variation in contribution amounts and timing ($t = 3.040, p=0.020$), or low variation in contribution amounts and timing ($t = 4.667, p=0.001$). We find no significant difference in SSI scores between any other pair-wise comparisons.

An omnibus one-way ANOVA also shows a significant difference in the baseline interest in the project category ($F=3.195, p=0.024$) across participants' selection groups. Yet, post-hoc Tukey's Honest Significant Difference (HSD) tests show no significant differences in pairwise comparisons except between participants that select a project at random and those that select a project for reasons other than the observed project descriptions or crowd signals ($t = 2.863, p=0.034$). The ANOVA shows no significant difference in the participants' altruism scores across the groups.

\begin{table}[!tp]
\caption{Study II: Post-hoc Tukey Honestly Significant Difference (HSD) test results for one-way ANOVA on participants' susceptibility to social influence across different reasons for selection according to open-ended responses. ``High'' denotes that the participant reports choosing the project and contribution list because of some aspect  of the heterogeneous amounts and timing. ``Low'' means that the participant mentions their selection being driven by the homogeneous amounts and timing. ``Random'' denotes a selection based on pure chance. P-values significant at: p<0.001***, p<0.01**, and p<0.05*.}
\label{tab:tukey_b}
\centering
\begin{tabular}{l|cccccc|}
\cline{1-7} %\cline{2-7}
\multicolumn{1}{|l|}{\textbf{Susceptibility to Social Influence}} & \multicolumn{1}{c|}{\begin{tabular}[c]{@{}c@{}}mean\\ (group1)\end{tabular}} & \multicolumn{1}{c|}{\begin{tabular}[c]{@{}c@{}}mean\\ (group2)\end{tabular}} & \multicolumn{1}{c|}{diff} & \multicolumn{1}{c|}{se} & \multicolumn{1}{c|}{T} & \multicolumn{1}{c|}{p-tukey} \\ \hline
\multicolumn{1}{|l|}{High - Low} & \multicolumn{1}{l|}{31.108} & \multicolumn{1}{l|}{37.000} & \multicolumn{1}{l|}{-5.892} & \multicolumn{1}{l|}{2.368} & \multicolumn{1}{l|}{-2.488} & 0.094 \\
\multicolumn{1}{|l|}{High - Random *} & \multicolumn{1}{l|}{31.108} & \multicolumn{1}{l|}{23.289} & \multicolumn{1}{l|}{7.818} & \multicolumn{1}{l|}{2.572} & \multicolumn{1}{l|}{3.040} & 0.020 \\
\multicolumn{1}{|l|}{Low - Random ***} & \multicolumn{1}{l|}{37.000} & \multicolumn{1}{l|}{23.289} & \multicolumn{1}{l|}{13.711} & \multicolumn{1}{l|}{2.938} & \multicolumn{1}{l|}{4.667} & 0.001 \\
\multicolumn{1}{|l|}{Description - High} & \multicolumn{1}{l|}{31.318} & \multicolumn{1}{l|}{31.108} & \multicolumn{1}{l|}{0.210} & \multicolumn{1}{l|}{1.554} & \multicolumn{1}{l|}{0.135} & 0.900 \\
\multicolumn{1}{|l|}{Description - Low} & \multicolumn{1}{l|}{31.318} & \multicolumn{1}{l|}{37.000} & \multicolumn{1}{l|}{-5.682} & \multicolumn{1}{l|}{2.105} & \multicolumn{1}{l|}{-2.699} & 0.055 \\
\multicolumn{1}{|l|}{Description - Random **} & \multicolumn{1}{l|}{31.318} & \multicolumn{1}{l|}{23.289} & \multicolumn{1}{l|}{8.028} & \multicolumn{1}{l|}{2.332} & \multicolumn{1}{l|}{3.443} & 0.005 \\
\hline
\end{tabular}
\end{table}

\subsection{Summary of Study II}
Study II demonstrates the efficiency of signaling in a more natural context than Study I and, surprisingly, it highlights just how unaware people are of this signaling in a crowdfunding setting. Participants systematically choose projects with high variance in contribution amounts and timing, for the most part attributing their selection to details in the description, such as higher need or better arguments. Our qualitative analysis reinforces this finding. The control condition shows uniform preferences for project descriptions across a variety of topics, fundraising goals, description lengths and quality. These findings are robust even with thorough counterbalancing and across differences in participants' interest levels in the projects and their altruistic attitudes. Those who score higher on susceptibility to social influence, are more responsive to signaling. Altogether, Study II indicates that the signaling encoded in contribution lists via heterogeneous or homogeneous amounts and timing is robust across various stimuli, conditions, and realistic contexts.

\section{Discussion}
\label{sec:discussion}

In this paper, we shed light on critical questions regarding social signaling on crowdfunding platforms: \emph{How, when, and why are crowd signals linked to potential contributors' decision-making?} We offer the first systematic large-scale experiments that investigate whether and under what conditions crowd signals influence participants' decisions to contribute to crowdfunding projects. Observing the choices and reflections of former users of crowdfunding platforms who have varying levels of susceptibility to social influence and altruistic attitudes, we present nuanced evidence that extends prior observational studies. Our evidence provides new insights about the fundamental ways that potential contributors perceive social signaling via prior contributions.

Specifically, our research validates previous findings about the role of non-trivial crowd signals in determining fundraising success~\cite{dambanemuya2019harnessing,dambanemuya2021multi,horvat2023hidden}. The novel experimental and qualitative evidence substantially refines our understanding of \emph{how} effective social signaling is even when potential contributors do not realize the impact crowd signals have on their decisions. Our large participant sample also provides essential novel insights about \emph{when} crowd signals work, highlighting that they are most salient among participants who are more susceptible to social influence. Participants' open-ended responses further enable us to uncover \emph{why} they respond differently to the observed crowd signals. Our discoveries establish the foundations for further theoretical work on mechanisms of social signaling in and beyond crowdfunding.

We argue that the confluence of individual decision-making, social signaling between crowd members, and emergent group behavior is essential for the attractiveness and success of crowdfunding. Crowdfunding platforms provide more than financial transactions. They also satisfy people's social and cognitive needs~\cite{gerber_crowdfunding_2013}. For that reason, contributors' autonomy in choosing meritorious projects, their opportunity to learn from others about project quality, and their experience of being part of successful collectives are indispensable for positive perceptions about crowdfunding and the sustainability of this crucial form of crowdsourcing. We argue that our work makes important findings at this critical intersection.

\subsection{Theoretical Implications}

Our work provides new insights for recent discussions about the effectiveness of social translucence in crowdfunding~\cite{kim2020enriched}. Although funders send signals via their observable actions on crowdfunding platforms, potential contributors are not always cognizant of how this signaling factors into their decision-making and complements observable project attributes. Our first study tested the fundamental signaling mechanism in a setting that primed participants to observe some role of crowd signals. Our second study clearly showed that even though participants still overwhelmingly selected high variance in contribution amounts and timing, they attributed their choices to the higher needs or opportunities that they perceived from the project descriptions. Thus, Study II demonstrates that simply making social information visible on crowdfunding sites is insufficient to enable social translucence. To be useful, social translucence theory requires an awareness of the signals encoded in the social information~\cite{erickson2000social}, which we observed lacking in more than half of the participants, who report having selected projects solely based on the description.

On a more basic level, our findings also contribute to ongoing debates on whether the visibility of others' contributions in crowdfunding triggers further engagement. For example, while some studies suggest that the visibility of such information increases contribution amounts~\cite{andreoni2004public,soetevent2005anonymity,van2013national}, other studies find a negative relationship between initial contributions and final success if those first amounts were low~\cite{koning2013experimental}. Our results point to an interesting possibility that warrants further research. Could the specific signals encoded in the social information matter more than the mere presence of social information? Our work has shown that crowd signals better explain the contexts where social information leads to increased contributions.

Unpacking the link between crowd signals and individual decision-making, we examined participants' open-ended responses about why they chose one contribution list and/or project description over another. We found compelling qualitative evidence for why high variation in contribution amounts and timing are associated with successful fundraising in observational studies. While existing studies emphasize the importance of large contribution amounts in signaling funders' confidence in a project's potential success~\cite{colombo2015internal,koning2013experimental,vulkan2016equity}, our findings suggest that small amounts make contributors with modest pledges more comfortable and confident to contribute. In other words, showing a variety of small and large contributions can be more beneficial for crowdfunding campaigns than simply showing large contributions, as the range increases the project's appeal to a more diverse pool of potential funders.

Consistent with signaling theory, participants in our study interpret the crowd signals differently. Our qualitative analysis of participants' reasons for selection shows that even among participants that select the same treatment condition, the reasons behind the selection might be different. For example, some participants interpret the high variation crowd signals (heterogeneous contribution amounts and timing) as signaling a project's broad appeal. Others interpret the same high variation signals as an opportunity to contribute at any level of giving. Still others perceive the small amounts as signaling a project's struggle to raise funds. Conversely, some participants interpret the low variation crowd signals (homogeneous contribution amounts and timing) as a projection of generosity, while others take the same signals as consensus among funders about a project's merit. Others think that the project's outcome is more predictable due to the consistency in contribution amounts and timing. Overall, we observe that most participants choose consistently high over low variation in contribution amounts and times. In both studies, this tendency is remarkably robust to different project categories, target amounts, description length and quality. By demonstrating the consistent role of these signals in project selection, our work lays the groundwork for theory-building in this area of collective decision-making.

\subsection{Practical Implications for Crowdfunding Platform Designers and Users}

Our findings have implications for all three major stakeholders involved in crowdfunding: the platforms, project creators, and potential funders.

Platform maintainers can build on our results to develop crowdfunding sites that harness crowd signals to improve social translucence, information acquisition by potential contributors, and resource allocation to meritorious projects. This will ensure the long-term success of their service. For example, we show that non-trivial differences in what contribution timing and amounts  users see can significantly affect fundraising success. Designers should thus be intentional about how many contributions they display and how visible they make the amounts and the arrival times of funds. Our findings suggest that these design choices are essential. Platforms can build on this new knowledge to devise ways to support signaling and coordination between funders.

Project creators can also use our results to improve the success of their campaigns. In alignment with prior work that demonstrates the importance of mobilizing a community in crowdfunding~\cite{hui2014understanding}, our findings indicate that project creators should diversify their outreach efforts to multiple funder categories. In particular, they should not only target the relatively ``big funders'' to grow the expected level of capital in-flow but should also reach out to ``small contributors'' to increase the fundraising effort's public appeal. As we demonstrate in our paper, showing high variation in contribution amounts and times can signal a project's broad appeal across various contributor groups, including those that can only contribute small amounts. This, in turn, could make fundraising feel more like an authentic community effort. Such endeavors may increase a project's chances of success and enhance project creators' effectiveness on crowdfunding platforms.

Finally, our main finding about the essential role of social signaling between prior and potential contributors has important consequences for funders. Our experiments expose the impact of specific crowd signals notwithstanding contributors' ignorance of their inherent reaction to such signaling during decision-making. This crucial insight calls for better educating platform users on both the positive and negative effects of signaling and herding~\cite{dambanemuya2023herding}. More broadly, we argue that such information should represent a fundamental part of digital literacy education efforts in general~\cite{van2017compoundness,malik2016identifying,hargittai2019internet}.

\subsection{Limitations and Future Work} 
Our study has a few limitations we hope future work can address effectively. First, this study investigates only one operationalization of crowd signals associated with successful fundraising. While we rely on prior empirical work on different platforms to create high and low variation conditions linked to funding outcomes~\cite{dambanemuya2021multi,dambanemuya2019harnessing,horvat2023hidden}, we cannot exclude the possibility that other crowd signals are also worth testing experimentally. The experimental setup we introduce here can serve as a fundamental approach for new investigations when further robust measurements of crowd signaling are discovered.

Second, we study only the effect of the crowd signals toward the end of the crowdfunding period (i.e., when 80\% of the funds have been raised). It is unclear to what extent this subtle indication that the fundraising campaign is nearing completion influences participants. Further work should investigate how crowd signals influence participant behavior when their contributions are solicited, e.g., at the beginning of the fundraising effort. 

Third, it is essential to highlight that participants in our study did not spend their own money. While one can envision experiments that request participants to spend research funds, these would be difficult if larger amounts (e.g., \$250) were involved. Large contributions represent the necessary counterpoint to small amounts in the high variation condition. Hence, if the goal was to scrutinize to what extent involving real money affects participants' decisions, one would likely need to compromise on the scale of the studies.

\subsection{Conclusion}
Our research focuses on how signaling among funders affects their choices and the success of funding requests. We validate the ability to predict fundraising success from crowd signals that quantify variation in contribution amounts and times. We show with two experimental studies that participants consistently select contribution lists with large variations in contribution amounts and timing, even when they attribute their choice to nonexistent differences in project descriptions. Uncovering the link between crowd signals and individual decision-making, we demonstrate that the signals are robust to participants' susceptibility to social influence, altruistic tendencies, and baseline interest in various project categories. Our results provide not only novel insights into an essential issue in online capital allocation but also an open problem in understanding the link between mechanisms of social influence and success on online platforms. We hope our work will improve the efficiency of crowdfunding and advance future research on the mechanisms of social signaling, even beyond the context of crowdfunding.

\section*{Acknowledgments} 
This work was supported by the U.S. National Science Foundation under Grant No. IIS-1755873 and a Northwestern University School of Communication COVID-19 Recovery Grant.

%%
%% The next two lines define the bibliography style to be used, and
%% the bibliography file.
\bibliographystyle{unsrt}  
\bibliography{main} 

\begin{thebibliography}{100}

\bibitem{mollick2014dynamics}
Ethan Mollick.
\newblock The dynamics of crowdfunding: {A}n exploratory study.
\newblock {\em Journal of Business Venturing}, 29(1):1--16, 2014.

\bibitem{vachelard2016guide}
Julien Vachelard, Thaise Gambarra-Soares, Gabriela Augustini, Pablo Riul, and
  Vinicius Maracaja-Coutinho.
\newblock A guide to scientific crowdfunding.
\newblock {\em PLoS Biology}, 14(2):e1002373, 2016.

\bibitem{kuo2012design}
Pei-Yi Kuo and Elizabeth Gerber.
\newblock Design principles: {C}rowdfunding as a creativity support tool.
\newblock In {\em CHI'12 Extended Abstracts on Human Factors in Computing
  Systems}, pages 1601--1606. 2012.

\bibitem{vulkan2016equity}
Nir Vulkan, Thomas Astebro, and Manuel~Fernandez Sierra.
\newblock Equity crowdfunding: A new phenomena.
\newblock {\em Journal of Business Venturing Insights}, 5:37 -- 49, 2016.

\bibitem{lee2022new}
Sumin Lee and Vili Lehdonvirta.
\newblock New digital safety net or just more ‘friendfunding’?
  {I}nstitutional analysis of medical crowdfunding in the united states.
\newblock {\em Information, Communication \& Society}, 25(8):1151--1175, 2022.

\bibitem{whitehouse2016}
Thomas Kalil and Rand Doug.
\newblock The promise of crowdfunding and american innovation, 2016.
\newblock Retrieved on October 20, 2022 from
  \url{https://obamawhitehouse.archives.gov/blog/2016/06/08/promise-crowdfunding-and-american-innovation}.

\bibitem{zhang2018bibliometric}
Wei Zhang, Yan-Chun Zhu, and Xiao-Lin Wu.
\newblock A bibliometric analysis of crowdfunding related research: Current
  trends and future prospect.
\newblock In {\em Proceedings of the 3rd International Conference on Crowd
  Science and Engineering}, pages 1--7, 2018.

\bibitem{greenberg_learning_2014}
Michael~D. Greenberg and Elizabeth~M. Gerber.
\newblock Learning to fail: {E}xperiencing public failure online through
  crowdfunding.
\newblock In {\em Proceedings of the SIGCHI Conference on Human Factors in
  Computing Systems}, {CHI} '14, pages 581--590, Toronto, Ontario, Canada,
  April 2014. Association for Computing Machinery.

\bibitem{crowddata2022}
The~Crowdfunding Center.
\newblock Platforms stats \& analytics: Showing period 1st {J}anuary 2014 -
  12th {J}anuary 2022.
\newblock {\em Retrieved from
  \url{https://www.thecrowdfundingcenter.com/data/platforms\#platforms_success}},
  2022.

\bibitem{kim2020enriched}
Jennifer~G Kim, Ha-Kyung Kong, Hwajung Hong, and Karrie Karahalios.
\newblock Enriched social translucence in medical crowdfunding.
\newblock In {\em Proceedings of the 2020 ACM Designing Interactive Systems
  Conference}, pages 1465--1477, 2020.

\bibitem{suh2008lifting}
Bongwon Suh, Ed~H Chi, Aniket Kittur, and Bryan~A Pendleton.
\newblock Lifting the veil: improving accountability and social transparency in
  wikipedia with wikidashboard.
\newblock In {\em Proceedings of the SIGCHI Conference on Human Factors in
  Computing Systems}, pages 1037--1040, 2008.

\bibitem{gilbert2012designing}
Eric Gilbert.
\newblock Designing social translucence over social networks.
\newblock In {\em Proceedings of the SIGCHI Conference on Human Factors in
  Computing Systems}, pages 2731--2740, 2012.

\bibitem{ehsan2021expanding}
Upol Ehsan, Q~Vera Liao, Michael Muller, Mark~O Riedl, and Justin~D Weisz.
\newblock Expanding explainability: Towards social transparency in ai systems.
\newblock In {\em Proceedings of the 2021 CHI Conference on Human Factors in
  Computing Systems}, pages 1--19, 2021.

\bibitem{erickson2000social}
Thomas Erickson and Wendy~A Kellogg.
\newblock Social translucence: {a}n approach to designing systems that support
  social processes.
\newblock {\em ACM Transactions on Computer-Human Interaction (TOCHI)},
  7(1):59--83, 2000.

\bibitem{erickson2002social}
Thomas Erickson, Christine Halverson, Wendy~A Kellogg, Mark Laff, and Tracee
  Wolf.
\newblock Social translucence: designing social infrastructures that make
  collective activity visible.
\newblock {\em Communications of the ACM}, 45(4):40--44, 2002.

\bibitem{larrimore2011peer}
Laura Larrimore, Li~Jiang, Jeff Larrimore, David Markowitz, and Scott Gorski.
\newblock Peer to peer lending: {T}he relationship between language features,
  trustworthiness, and persuasion success.
\newblock {\em Journal of Applied Communication Research}, 39(1):19--37, 2011.

\bibitem{althoff2014ask}
Tim Althoff, Cristian Danescu-Niculescu-Mizil, and Dan Jurafsky.
\newblock How to ask for a favor: {A} case study on the success of altruistic
  requests.
\newblock In {\em Eighth International AAAI Conference on Weblogs and Social
  Media}, 2014.

\bibitem{mitra2014language}
Tanushree Mitra and Eric Gilbert.
\newblock The language that gets people to give: Phrases that predict success
  on kickstarter.
\newblock In {\em Proceedings of the 17th ACM Conference on Computer Supported
  Cooperative Work \& Social Computing}, pages 49--61, 2014.

\bibitem{rhue_emotional_2018}
Lauren Rhue and Lionel~P. Robert.
\newblock Emotional delivery in pro-social crowdfunding success.
\newblock In {\em Extended Abstracts of the 2018 CHI Conference on Human
  Factors in Computing Systems}, {CHI} {EA} '18, pages 1--6, Montreal QC,
  Canada, April 2018. Association for Computing Machinery.

\bibitem{gafni2019life}
Hadar Gafni, Dan Marom, and Orly Sade.
\newblock Are the life and death of an early-stage venture indeed in the power
  of the tongue? {L}essons from online crowdfunding pitches.
\newblock {\em Strategic Entrepreneurship Journal}, 13(1):3--23, 2019.

\bibitem{anglin2018power}
Aaron~H Anglin, Jeremy~C Short, Will Drover, Regan~M Stevenson, Aaron~F
  McKenny, and Thomas~H Allison.
\newblock The power of positivity? {T}he influence of positive psychological
  capital language on crowdfunding performance.
\newblock {\em Journal of Business Venturing}, 33(4):470--492, 2018.

\bibitem{colombo2015internal}
Massimo~G Colombo, Chiara Franzoni, and Cristina Rossi-Lamastra.
\newblock Internal social capital and the attraction of early contributions in
  crowdfunding.
\newblock {\em Entrepreneurship Theory and Practice}, 39(1):75--100, 2015.

\bibitem{greiner2009role}
Martina~E Greiner and Hui Wang.
\newblock The role of social capital in people-to-people lending marketplaces.
\newblock {\em ICIS 2009 Proceedings}, page~29, 2009.

\bibitem{zheng2014role}
Haichao Zheng, Dahui Li, Jing Wu, and Yun Xu.
\newblock The role of multidimensional social capital in crowdfunding: {A}
  comparative study in china and us.
\newblock {\em Information \& Management}, 51(4):488--496, 2014.

\bibitem{xu2014show}
Anbang Xu, Xiao Yang, Huaming Rao, Wai-Tat Fu, Shih-Wen Huang, and Brian~P
  Bailey.
\newblock Show me the money! {A}n analysis of project updates during
  crowdfunding campaigns.
\newblock In {\em Proceedings of the SIGCHI Conference on Human Factors in
  Computing Systems}, pages 591--600, 2014.

\bibitem{dey2017art}
Sanorita Dey, Brittany Duff, Karrie Karahalios, and Wai-Tat Fu.
\newblock The art and science of persuasion: {n}ot all crowdfunding campaign
  videos are the same.
\newblock In {\em Proceedings of the 2017 ACM Conference on Computer Supported
  Cooperative Work and Social Computing}, pages 755--769, 2017.

\bibitem{shang2009field}
Jen Shang and Rachel Croson.
\newblock A field experiment in charitable contribution: The impact of social
  information on the voluntary provision of public goods.
\newblock {\em The Economic Journal}, 119(540):1422--1439, 2009.

\bibitem{zhang2012rational}
Juanjuan Zhang and Peng Liu.
\newblock Rational herding in microloan markets.
\newblock {\em Management Science}, 58(5):892--912, 2012.

\bibitem{ceyhan2011dynamics}
Simla Ceyhan, Xiaolin Shi, and Jure Leskovec.
\newblock Dynamics of bidding in a {P2P} lending service: {E}ffects of herding
  and predicting loan success.
\newblock In {\em Proceedings of the 20th International Conference on World
  Wide Web}, pages 547--556. ACM, 2011.

\bibitem{burtch2013empirical}
Gordon Burtch, Anindya Ghose, and Sunil Wattal.
\newblock An empirical examination of the antecedents and consequences of
  contribution patterns in crowd-funded markets.
\newblock {\em Information Systems Research}, 24(3):499--519, 2013.

\bibitem{dambanemuya2019harnessing}
Henry~K Dambanemuya and Em{\H{o}}ke-{\'A}gnes Horv{\'a}t.
\newblock Harnessing collective intelligence in {P2P} lending.
\newblock In {\em Proceedings of the 10th ACM Conference on Web Science}, pages
  57--64, 2019.

\bibitem{van2020follow}
Claire van Teunenbroek, Ren{\'e} Bekkers, et~al.
\newblock Follow the crowd: Social information and crowdfunding donations in a
  large field experiment.
\newblock {\em Journal of Behavioral Public Administration}, 3(1), 2020.

\bibitem{horvat2023hidden}
Em\H{o}ke-\'{A}gnes Horv\'{a}t, Henry~Kudzanai Dambanemuya, Jayaram Uparna, and
  Brian Uzzi.
\newblock Hidden indicators of collective intelligence in crowdfunding.
\newblock In {\em Proceedings of the ACM Web Conference 2023}, WWW '23, page
  3806–3815, New York, NY, USA, 2023. Association for Computing Machinery.

\bibitem{spence2002signaling}
Michael Spence.
\newblock Signaling in retrospect and the informational structure of markets.
\newblock {\em American Economic Review}, 92(3):434--459, 2002.

\bibitem{salganik2006experimental}
Matthew~J Salganik, Peter~Sheridan Dodds, and Duncan~J Watts.
\newblock Experimental study of inequality and unpredictability in an
  artificial cultural market.
\newblock {\em Science}, 311(5762):854--856, 2006.

\bibitem{connelly2011signaling}
Brian~L Connelly, S~Trevis Certo, R~Duane Ireland, and Christopher~R Reutzel.
\newblock Signaling theory: A review and assessment.
\newblock {\em Journal of Management}, 37(1):39--67, 2011.

\bibitem{vismara2018information}
Silvio Vismara.
\newblock Information cascades among investors in equity crowdfunding.
\newblock {\em Entrepreneurship Theory and Practice}, 42(3):467--497, 2018.

\bibitem{astebro2019herding}
Thomas~B Astebro, Manuel Fern{\'a}ndez~Sierra, Stefano Lovo, and Nir Vulkan.
\newblock Herding in equity crowdfunding.
\newblock In {\em Finance Meeting Eurofidai-AFFI, Paris}, 2019.

\bibitem{zaggl2019small}
Michael~A Zaggl and Joern Block.
\newblock Do small funding amounts lead to reverse herding? {A} field
  experiment in reward-based crowdfunding.
\newblock {\em Journal of Business Venturing Insights}, 12:e00139, 2019.

\bibitem{chan2020bellwether}
CS~Richard Chan, Annaleena Parhankangas, Arvin Sahaym, and Pyayt Oo.
\newblock Bellwether and the herd? {U}npacking the u-shaped relationship
  between prior funding and subsequent contributions in reward-based
  crowdfunding.
\newblock {\em Journal of Business Venturing}, 35(2):105934, 2020.

\bibitem{zakhlebin2019investor}
Igor Zakhlebin and Em{\H{o}}ke-{\'A}gnes Horv{\'a}t.
\newblock Investor retention in equity crowdfunding.
\newblock In {\em Proceedings of the 10th ACM Conference on Web Science}, pages
  343--351, 2019.

\bibitem{dambanemuya2023herding}
Henry~Kudzanai Dambanemuya, Johannes Wachs, and Em{\H o}ke-{\'A}gnes Horv\'at.
\newblock Understanding (ir)rational herding online.
\newblock In {\em Proceedings of The ACM Collective Intelligence Conference},
  CI '23, page 79–88, New York, NY, USA, 2023. Association for Computing
  Machinery.

\bibitem{dambanemuya2021multi}
Henry~K Dambanemuya and Em{\H{o}}ke-{\'A}gnes Horv{\'a}t.
\newblock A multi-platform study of crowd signals associated with successful
  online fundraising.
\newblock {\em Proceedings of the ACM on Human-Computer Interaction},
  5(CSCW1):1--19, 2021.

\bibitem{chakraborty2019impact}
Abhijnan Chakraborty, Nuno Mota, Asia~J Biega, Krishna~P Gummadi, and Hoda
  Heidari.
\newblock On the impact of choice architectures on inequality in online
  donation platforms.
\newblock In {\em The World Wide Web Conference}, pages 2623--2629, 2019.

\bibitem{thalerchoice}
Richard~H Thaler, Cass~R Sunstein, and John~P Balz.
\newblock Choice architecture, 2014.
\newblock {\em The Behavioral Foundations of Public Policy}, pages 428--439.

\bibitem{aragon2017detecting}
Pablo Arag{\'o}n, Vicen{\c{c}} G{\'o}mez, and Andreas Kaltenbrunner.
\newblock Detecting platform effects in online discussions.
\newblock {\em Policy \& Internet}, 9(4):420--443, 2017.

\bibitem{lee_content-based_2018}
SeungHun Lee, KangHee Lee, and Hyun-chul Kim.
\newblock Content-based {Success} {Prediction} of {Crowdfunding} {Campaigns}:
  {A} {Deep} {Learning} {Approach}.
\newblock In {\em Companion of the 2018 {ACM} {Conference} on {Computer}
  {Supported} {Cooperative} {Work} and {Social} {Computing}}, {CSCW} '18, pages
  193--196, Jersey City, NJ, USA, October 2018. Association for Computing
  Machinery.

\bibitem{etter2013launch}
Vincent Etter, Matthias Grossglauser, and Patrick Thiran.
\newblock Launch hard or go home! {P}redicting the success of kickstarter
  campaigns.
\newblock In {\em Proceedings of the first ACM Conference on Online Social
  Networks}, pages 177--182, 2013.

\bibitem{lu2014inferring}
Chun-Ta Lu, Sihong Xie, Xiangnan Kong, and Philip~S Yu.
\newblock Inferring the impacts of social media on crowdfunding.
\newblock In {\em Proceedings of the 7th ACM International Conference on Web
  Search and Data Mining}, pages 573--582, 2014.

\bibitem{zhang_predicting_2017}
Qizhen Zhang, Tengyuan Ye, Meryem Essaidi, Shivani Agarwal, Vincent Liu, and
  Boon~Thau Loo.
\newblock Predicting startup crowdfunding success through longitudinal social
  engagement analysis.
\newblock In {\em Proceedings of the 2017 {ACM} on {Conference} on
  {Information} and {Knowledge} {Management}}, {CIKM} '17, pages 1937--1946,
  Singapore, Singapore, November 2017. Association for Computing Machinery.

\bibitem{cumming2015crowdfunding}
Douglas~J Cumming, Ga{\"e}l Leboeuf, and Armin Schwienbacher.
\newblock Crowdfunding models: Keep-it-all vs. all-or-nothing.
\newblock {\em Financial Management}, 2015.

\bibitem{cordova2015determinants}
Alessandro Cordova, Johanna Dolci, and Gianfranco Gianfrate.
\newblock The determinants of crowdfunding success: {E}vidence from technology
  projects.
\newblock {\em Procedia-Social and Behavioral Sciences}, 181:115--124, 2015.

\bibitem{collier2010sending}
Benjamin~C Collier and Robert Hampshire.
\newblock Sending mixed signals: {M}ultilevel reputation effects in
  peer-to-peer lending markets.
\newblock In {\em Proceedings of the 2010 ACM Conference on Computer Supported
  Cooperative Work}, pages 197--206, 2010.

\bibitem{ahlers2015signaling}
Gerrit~KC Ahlers, Douglas Cumming, Christina G{\"u}nther, and Denis Schweizer.
\newblock Signaling in equity crowdfunding.
\newblock {\em Entrepreneurship Theory and Practice}, 39(4):955--980, 2015.

\bibitem{horvat2015network}
Em{\H{o}}ke-{\'A}gnes Horv{\'a}t, Jayaram Uparna, and Brian Uzzi.
\newblock Network vs market relations: The effect of friends in crowdfunding.
\newblock In {\em Proceedings of the 2015 IEEE/ACM International Conference on
  Advances in Social Networks Analysis and Mining 2015}, pages 226--233, 2015.

\bibitem{vismara2016equity}
Silvio Vismara.
\newblock Equity retention and social network theory in equity crowdfunding.
\newblock {\em Small Business Economics}, 46(4):579--590, 2016.

\bibitem{chung2015long}
Jinwook Chung and Kyumin Lee.
\newblock A long-term study of a crowdfunding platform: Predicting project
  success and fundraising amount.
\newblock In {\em Proceedings of the 26th ACM Conference on Hypertext \& Social
  Media}, pages 211--220, 2015.

\bibitem{koning2013experimental}
Rembrand Koning and Jacob Model.
\newblock Experimental study of crowdfunding cascades: {W}hen nothing is better
  than something.
\newblock {\em Available at SSRN 2308161}, 2013.

\bibitem{solomon2015don}
Jacob Solomon, Wenjuan Ma, and Rick Wash.
\newblock Don't wait! {H}ow timing affects coordination of crowdfunding
  donations.
\newblock In {\em Proceedings of the 18th ACM Conference on Computer Supported
  Cooperative Work \& Social Computing}, pages 547--556, 2015.

\bibitem{agrawal2015crowdfunding}
Ajay Agrawal, Christian Catalini, and Avi Goldfarb.
\newblock Crowdfunding: {G}eography, social networks, and the timing of
  investment decisions.
\newblock {\em Journal of Economics \& Management Strategy}, 24(2):253--274,
  2015.

\bibitem{vismara2016information}
Silvio Vismara.
\newblock Information cascades among investors in equity crowdfunding.
\newblock {\em Entrepreneurship Theory and Practice}, 2016.

\bibitem{vismara2018signaling}
Silvio Vismara.
\newblock Signaling to overcome inefficiencies in crowdfunding markets.
\newblock In {\em The Economics of Crowdfunding}, pages 29--56. Springer, 2018.

\bibitem{dobbie2021measuring}
Will Dobbie, Andres Liberman, Daniel Paravisini, and Vikram Pathania.
\newblock Measuring bias in consumer lending.
\newblock {\em The Review of Economic Studies}, 88(6):2799--2832, 2021.

\bibitem{stevenson2019out}
Regan~M Stevenson, Michael~P Ciuchta, Chaim Letwin, Jenni~M Dinger, and
  Jeffrey~B Vancouver.
\newblock Out of control or right on the money? {F}under self-efficacy and
  crowd bias in equity crowdfunding.
\newblock {\em Journal of Business Venturing}, 34(2):348--367, 2019.

\bibitem{gerber_crowdfunding_2013}
Elizabeth~M. Gerber and Julie Hui.
\newblock Crowdfunding: {M}otivations and deterrents for participation.
\newblock {\em ACM Transactions on Computer-Human Interaction (TOCHI)},
  20(6):34:1--34:32, December 2013.

\bibitem{kraut2011encouraging}
Robert~E Kraut and Paul Resnick.
\newblock Encouraging contribution to online communities.
\newblock {\em Building successful online communities: Evidence-based social
  design}, pages 21--76, 2011.

\bibitem{kusumarani2019people}
Riri Kusumarani and Hangjung Zo.
\newblock Why people participate in online political crowdfunding: A civic
  voluntarism perspective.
\newblock {\em Telematics and Informatics}, 41:168--181, 2019.

\bibitem{steigenberger2017supporters}
Norbert Steigenberger.
\newblock Why supporters contribute to reward-based crowdfunding.
\newblock {\em International Journal of Entrepreneurial Behavior \& Research},
  2017.

\bibitem{giudici2018reward}
Giancarlo Giudici, Massimiliano Guerini, and Cristina Rossi-Lamastra.
\newblock Reward-based crowdfunding of entrepreneurial projects: {t}he effect
  of local altruism and localized social capital on proponents' success.
\newblock {\em Small Business Economics}, 50(2):307--324, 2018.

\bibitem{rodriguez2019altruism}
Yusimi Rodriguez-Ricardo, Maria Sicilia, and Manuela L{\'o}pez.
\newblock Altruism and internal locus of control as determinants of the
  intention to participate in crowdfunding: {T}he mediating role of trust.
\newblock {\em Journal of Theoretical and Applied Electronic Commerce
  Research}, 14(3):1--16, 2019.

\bibitem{hui2014understanding}
Julie~S Hui, Michael~D Greenberg, and Elizabeth~M Gerber.
\newblock Understanding the role of community in crowdfunding work.
\newblock In {\em Proceedings of the 17th ACM Conference on Computer Supported
  Cooperative Work \& Social Computing}, pages 62--74, 2014.

\bibitem{liu2021social}
Yang Liu, Yuan Chen, and Zhi-Ping Fan.
\newblock Do social network crowds help fundraising campaigns? {E}ffects of
  social influence on crowdfunding performance.
\newblock {\em Journal of Business Research}, 122:97--108, 2021.

\bibitem{kuppuswamy2018crowdfunding}
Venkat Kuppuswamy and Barry~L Bayus.
\newblock Crowdfunding creative ideas: The dynamics of project backers.
\newblock In {\em The Economics of Crowdfunding}, pages 151--182. Springer,
  2018.

\bibitem{spence1973job}
M~Spence.
\newblock Job market signaling.
\newblock {\em MIT Press}, 1973.

\bibitem{reichenbach2021signals}
Felix Reichenbach and Martin Walther.
\newblock Signals in equity-based crowdfunding and risk of failure.
\newblock {\em Financial Innovation}, 7(1):1--30, 2021.

\bibitem{plummer2016better}
Lawrence~A Plummer, Thomas~H Allison, and Brian~L Connelly.
\newblock Better together? {S}ignaling interactions in new venture pursuit of
  initial external capital.
\newblock {\em Academy of Management Journal}, 59(5):1585--1604, 2016.

\bibitem{fischbacher2001people}
Urs Fischbacher, Simon G{\"a}chter, and Ernst Fehr.
\newblock Are people conditionally cooperative? {E}vidence from a public goods
  experiment.
\newblock {\em Economics Letters}, 71(3):397--404, 2001.

\bibitem{bradley2019relatively}
Alex Bradley, Claire Lawrence, and Eamonn Ferguson.
\newblock When the relatively poor prosper: the underdog effect on charitable
  donations.
\newblock {\em Nonprofit and Voluntary Sector Quarterly}, 48(1):108--127, 2019.

\bibitem{zamudio2018e13}
C{\'e}sar Zamudio, Yiru Wang, et~al.
\newblock Rooting for {R}ocky or {A}pollo? {U}nderdog narratives and
  crowdfunding success.
\newblock {\em ACR North American Advances}, 2018.

\bibitem{duncan2004theory}
Brian Duncan.
\newblock A theory of impact philanthropy.
\newblock {\em Journal of Public Economics}, 88(9-10):2159--2180, 2004.

\bibitem{flesch2007flesch}
Rudolf Flesch.
\newblock Flesch-kincaid readability test.
\newblock {\em Retrieved October}, 26(3):2007, 2007.

\bibitem{rushton1981altruistic}
J~Philippe Rushton, Roland~D Chrisjohn, and G~Cynthia Fekken.
\newblock The altruistic personality and the self-report altruism scale.
\newblock {\em Personality and Individual Differences}, 2(4):293--302, 1981.

\bibitem{stockli2020susceptibility}
Sabrina St{\"o}ckli and Doris Hofer.
\newblock Susceptibility to social influence predicts behavior on facebook.
\newblock {\em PloS One}, 15(3):e0229337, 2020.

\bibitem{bearden1989measurement}
William~O Bearden, Richard~G Netemeyer, and Jesse~E Teel.
\newblock Measurement of consumer susceptibility to interpersonal influence.
\newblock {\em Journal of Consumer Research}, 15(4):473--481, 1989.

\bibitem{reynolds1971mutually}
Fred~D Reynolds and William~R Darden.
\newblock Mutually adaptive effects of interpersonal communication.
\newblock {\em Journal of Marketing Research}, 8(4):449--454, 1971.

\bibitem{hargittai2020comparing}
Eszter Hargittai and Aaron Shaw.
\newblock Comparing internet experiences and prosociality in {A}mazon
  {M}echanical {T}urk and population-based survey samples.
\newblock {\em Socius}, 6:2378023119889834, 2020.

\bibitem{shaw2021online}
Aaron Shaw and Eszter Hargittai.
\newblock Do the online activities of {A}mazon {M}echanical {T}urk workers
  mirror those of the general population? {A} comparison of two survey samples.
\newblock {\em International Journal of Communication}, 15:16, 2021.

\bibitem{morejon2016crowdfunding}
Roy Morejon.
\newblock Crowdfunding demographics and {K}ickstarter project statistics.
\newblock Art of the Kickstart Blog, 2016.
\newblock
  https://artofthekickstart.com/crowdfunding-demographics-kickstarter-project-statistics/.

\bibitem{berinsky2012evaluating}
Adam~J Berinsky, Gregory~A Huber, and Gabriel~S Lenz.
\newblock Evaluating online labor markets for experimental research:
  {A}mazon.com's {M}echanical {T}urk.
\newblock {\em Political Analysis}, 20(3):351--368, 2012.

\bibitem{goodman2013data}
Joseph~K Goodman, Cynthia~E Cryder, and Amar Cheema.
\newblock Data collection in a flat world: The strengths and weaknesses of
  mechanical turk samples.
\newblock {\em Journal of Behavioral Decision Making}, 26(3):213--224, 2013.

\bibitem{paolacci2010running}
Gabriele Paolacci, Jesse Chandler, and Panagiotis~G Ipeirotis.
\newblock Running experiments on {Am}azon {M}echanical {T}urk.
\newblock {\em Judgment and Decision Making}, 5(5):411--419, 2010.

\bibitem{peer2014reputation}
Eyal Peer, Joachim Vosgerau, and Alessandro Acquisti.
\newblock Reputation as a sufficient condition for data quality on {A}mazon
  {M}echanical {T}urk.
\newblock {\em Behavior Research Methods}, 46(4):1023--1031, 2014.

\bibitem{braun2006using}
Virginia Braun and Victoria Clarke.
\newblock Using thematic analysis in psychology.
\newblock {\em Qualitative Research in Psychology}, 3(2):77--101, 2006.

\bibitem{paas2018instructional}
Leonard~J Paas, Sara Dolnicar, and Logi Karlsson.
\newblock Instructional manipulation checks: A longitudinal analysis with
  implications for mturk.
\newblock {\em International Journal of Research in Marketing}, 35(2):258--269,
  2018.

\bibitem{hauser2016attentive}
David~J Hauser and Norbert Schwarz.
\newblock Attentive turkers: Mturk participants perform better on online
  attention checks than do subject pool participants.
\newblock {\em Behavior Research Methods}, 48(1):400--407, 2016.

\bibitem{oppenheimer2009instructional}
Daniel~M Oppenheimer, Tom Meyvis, and Nicolas Davidenko.
\newblock Instructional manipulation checks: {D}etecting satisficing to
  increase statistical power.
\newblock {\em Journal of Experimental Social Psychology}, 45(4):867--872,
  2009.

\bibitem{xu2022implications}
Lei Xu, Dahui Li, Chun-Hung Chiu, Qing Zhang, and Runpeng Gao.
\newblock Implications of warm-glow effect and risk aversion in reward-based
  crowdfunding.
\newblock {\em Transportation Research Part E: Logistics and Transportation
  Review}, 160:102681, 2022.

\bibitem{andreoni2004public}
James Andreoni and Ragan Petrie.
\newblock Public goods experiments without confidentiality: {A} glimpse into
  fund-raising.
\newblock {\em Journal of Public Economics}, 88(7-8):1605--1623, 2004.

\bibitem{soetevent2005anonymity}
Adriaan~R Soetevent.
\newblock Anonymity in giving in a natural context—a field experiment in 30
  churches.
\newblock {\em Journal of Public Economics}, 89(11-12):2301--2323, 2005.

\bibitem{van2013national}
Marco~HD Van~Leeuwen and Pamala Wiepking.
\newblock National campaigns for charitable causes: A literature review.
\newblock {\em Nonprofit and Voluntary Sector Quarterly}, 42(2):219--240, 2013.

\bibitem{van2017compoundness}
Alexander~JAM Van~Deursen, Ellen Helsper, Rebecca Eynon, and Jan~AGM Van~Dijk.
\newblock The compoundness and sequentiality of digital inequality.
\newblock {\em International Journal of Communication}, 11:452--473, 2017.

\bibitem{malik2016identifying}
Momin~M Malik and J{\"u}rgen Pfeffer.
\newblock Identifying platform effects in social media data.
\newblock In {\em Tenth International AAAI Conference on Web and Social Media},
  2016.

\bibitem{hargittai2019internet}
Eszter Hargittai and Marina Micheli.
\newblock Internet skills and why they matter.
\newblock {\em Society and the internet: How networks of information and
  communication are changing our lives}, 109, 2019.

\end{thebibliography}

%%
%% If your work has an appendix, this is the place to put it.
% \appendix

\newpage

\section*{Appendix}

\subsection*{Screening Questionnaire}
\label{sec:screening_qs}

The following questions were used to screen participants.

1. Which one of the following is TRUE about crowdfunding? 
\begin{enumerate}[label=(\alph*)]
    \item Crowdfunding is the practice of financing public and private green investments in environmental goods and services and prevention of damage to the environment.
    \item Crowdfunding is the practice of funding a project or venture by raising many small amounts of money from a large number of people, typically via online.
    \item Crowdfunding is an organized social movement to empower developing country producers and promoting sustainability.
    \item Crowdfunding is a peer-to-peer distributed ledger forged by consensus, combined with a system for smart contracts and other assistive technologies.
\end{enumerate}

2. Examples of crowdfunding include:
\begin{enumerate}[label=(\alph*)]
    \item Bing and google 
    \item Call a Bike and Uber 
    \item Kickstarter and Indiegogo
    \item WeWork and Regus 
\end{enumerate}

3. How have you participated crowdfunding? 
\begin{enumerate}[label=(\alph*)]
    \item Contributor
    \item Project Creator / Fundraiser
    \item All of the above
    \item Neither
\end{enumerate}

4. How often do you participate in crowdfunding?
\begin{enumerate}[label=(\alph*)]
    \item Daily
    \item At least once a week 
    \item At least once a month
    \item A few times per year 
    \item Never
\end{enumerate}

\end{document}